# Redox-tunable structural colour images based on UV-patterned conducting polymers

*Shangzhi Chen[1], Stefano Rossi[1], Ravi Shanker[1], Giancarlo Cincotti[1], Sampath Gamage[1], Philipp Kühne,[2,3], Vallery Stanishev[2], Isak Engquist[1], Magnus Berggren[1], Jesper Edberg[4], Vanya Darakchieva[2,3] and Magnus P. Jonsson[1]\**

[1] Laboratory of Organic Electronics, Department of Science and Technology (ITN), Linköping University, SE-601 74 Norrköping, Sweden
[2] Terahertz Materials Analysis Center (THeMAC), Department of Physics, Chemistry and Biology (IFM), Linköping University, SE-581 83 Linköping, Sweden
[3] Center for III-Nitride Technology, C3NiT-Janzèn, Department of Physics, Chemistry and Biology (IFM), Linköping University, SE-581 83 Linköping, Sweden
[4] RISE Research Institutes of Sweden, Bio- and Organic Electronics, Bredgatan 35, SE-602 21 Norrköping, Sweden
\* Correspondence: magnus.jonsson@liu.se

**Precise manipulation of light-matter interaction has enabled a wide variety of approaches to create bright and vivid structural colours. Techniques utilizing photonic crystals[1], Fabry-Pérot cavities[2], plasmonics[3-5], or high-refractive index dielectric metasurfaces[6,7] have been studied for applications ranging from optical coatings to reflective displays. However, complicated fabrication procedures for sub-wavelength nanostructures, limited active areas, and inherent absence of tunability with these approaches significantly impede their further developments towards flexible, large-scale, and switchable devices compatible with facile and cost-effective production. Herein, we present a way to generate structural colours based on conducting polymer thin films prepared on metallic surfaces *via* vapour phase polymerization and ultraviolet (UV) light patterning. Varying the UV dose leads to synergistic variation of film absorption and thickness, which generates controllable colours from violet to red. Together with greyscale photomasks this enables fabrication of high-resolution colour images using single exposure steps. We further demonstrate spatiotemporal tuning of the structurally coloured surfaces and images *via* electrochemical modulation of the polymer redox state. The simple structure, facile fabrication, wide colour gamut, and dynamic colour tuning make this concept competitive for future multi-functional and smart displays.**

Layered thin films structures, exemplified by Fabry-Pérot cavities[2,8,9] and ultrathin lossy films on metal mirrors[10], can produce reflective structural colours based on interference effects. They offer various attractive features, including intrinsic resistance to fading, excellent durability, bright colours under direct sunlight, and possibility for high resolution images[10-14]. Combined with means to switch or tune the structural colours, they show potential for full-colour electronic papers[15], with promise for ultralow power consumption, wide vibrant colour ranges, compact device structures, and high switching speeds[8,16]. Instead of defining structural colours only *via* geometrical factors (*i.e.,* film thickness)[2,17,18], our concept takes advantage of simultaneous modulation of both thickness and complex permittivity of UV-patterned conducting polymer thin films on metal mirrors (illustrated in **Figure 1a**). The conducting polymer properties and structural colouration are further dynamically tuneable via the polymers' intrinsic redox properties[19]. We demonstrate the concept using the conducting polymer poly(3,4-ethylenedioxythiophene):Tosylate (PEDOT:Tos, see chemical structure in **Figure 1b**) prepared by vapour phase polymerization (VPP). This method offers good and reproducible film quality[20-23] and also control of the polymer properties by UV light exposure of the oxidant precursors before polymerisation[24,25].

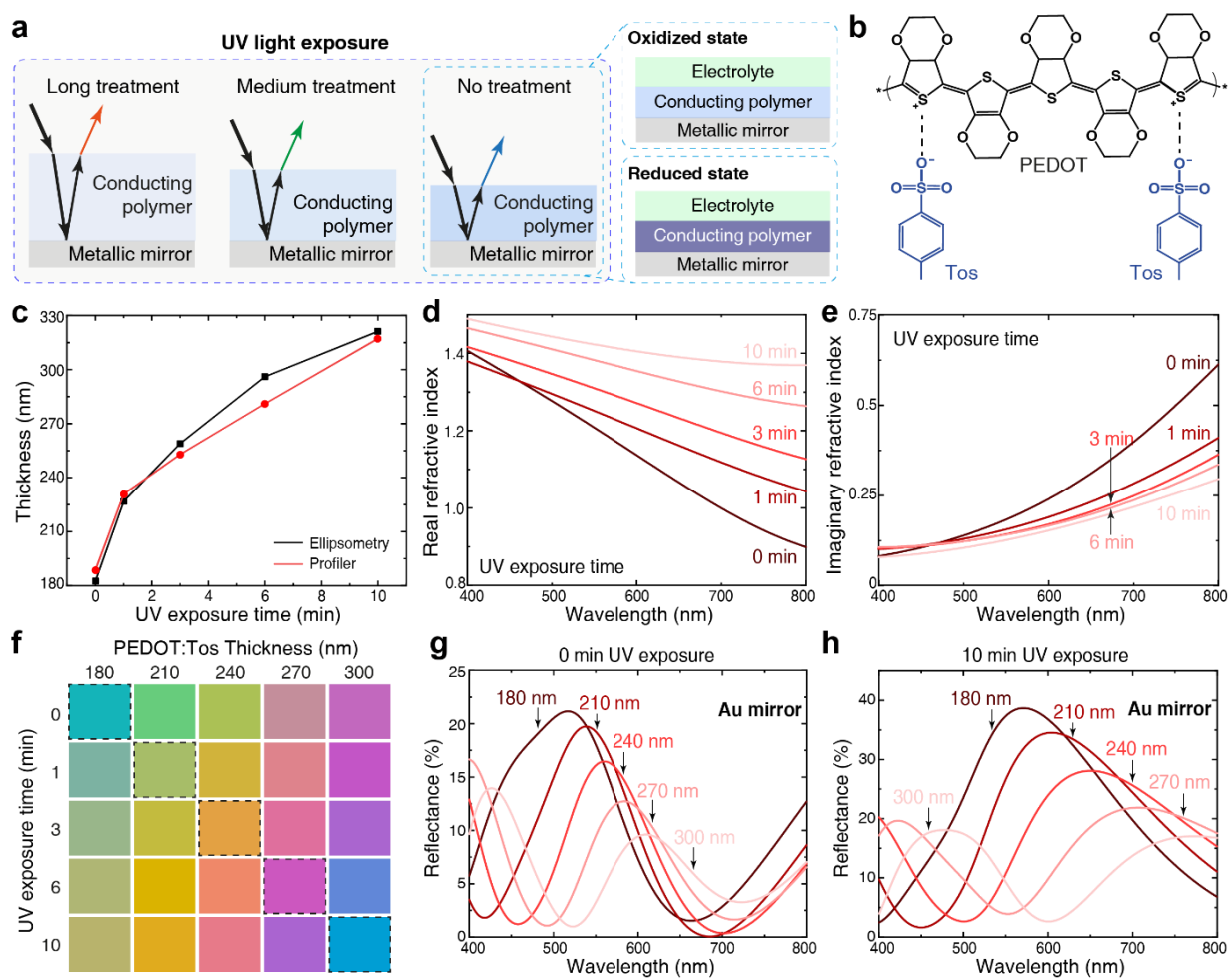

**Figure 1 | The concept of structural colours based on UV-treated conducting polymers. a,** Schematics of device structure, where by introducing different doze of UV light treatment both the thickness and the complex refractive index of the conducting polymer films are tuned to generate reflected light with different wavelengths. By depositing a layer of electrolyte on top, the reflected colours of the devices can be further tuned by controlling the conducting polymer redox state (reduced and oxidized states). **b,** Chemical structure of PEDOT:Tos. **c-e**, Thickness (**c**), in-plane real (**d**) and imaginary refractive index (**e**) changes of PEDOT:Tos films based on different UV exposure time (0, 1, 3, 6, and 10 min) and VPP time of 60 min. The thickness values were determined by both spectroscopic ellipsometry and profiler. **f,** Simulated pseudo-colours of devices colour palette by different UV exposure times and thicknesses. **g** and **h**, Simulated reflectance curves of the devices with 0 min (**g**) and 10 min (**h**) UV treatment. In the simulation, Au with thickness of 100 nm was used as the metallic mirror.

In accordance with previous studies[24], the thickness of our conducting polymer films could be altered in the range between 180 nm to 320 nm (**Figure 1c**) by varying only the UV exposure time of the spin-coated oxidant before VPP, and keeping other parameters fixed (see **Methods** for details). Ellipsometry measurements (**Figure S1** and **S2**) reveal that the thickness variation is accompanied by gradual variation in the complex refractive index of the conducting polymer films (**Figure 1d** and **1e**). We observe significant changes in the in-plane real refractive index in the red and near infrared (NIR) regions (see **Figure S2** for the out-of-plane data), with increasing values for longer UV exposure times. The imaginary component instead shows a significant decrease at long wavelengths upon UV exposure, and a tendency of slightly enhanced values at shorter wavelengths. These results agree with measured extinction spectra (**Figure S3**).

Numerical simulations predict that the conducting polymer films should be able to generate vibrant reflective structural colours if placed on a Au mirror (**Figure 1f**). The structural colours vary both with (UV-induced) permittivity modulation and polymer thickness. Using the UV-exposure approach, we can simultaneously modulate both those properties, which corresponds to the diagonal of the pseudo-colour palette presented in **Figure 1f**. Such simultaneous modulation leads to improved colour tuning due to several synergistic effects. First, the reflectance peak is redshifted not only by the increasing thickness, but also by the UV-induced increase in real permittivity. As a result, long-wavelength resonances can be generated at lower thicknesses (**Figure 1g** and **1h**), which leads to reduced absorption losses. At the same time, the large reduction in imaginary permittivity upon UV light exposure suppresses absorption at long wavelengths, which further aids the generation of yellow and red colours upon UV-treatment. **Figure 1h** also shows that the bluish structural colour for the longest UV exposure time (bottom right in **Figure 1f**) is due to a $2^{nd}$ order reflectance peak emerging at shorter wavelengths. Again, the UV exposure reduces the influence of this feature compared with only varying the thickness, both by allowing higher reflection for the primary red peak and also by increasing the spectral distance between the $1^{st}$ and $2^{nd}$ order peaks by reducing the negative slope of the real permittivity (also see **Figure S6 and S7**). Altogether, UV-induced tuning of the conducting polymer film provides several effects that work together to improve the colour gamut of the system.

We experimentally prepared PEDOT:Tos films on Au mirrors using different UV exposure times (0, 1, 3, 6, and 10 min) of the oxidant before VPP (**Figure 2a**). The resulting reflectance spectra are presented in **Figure 2b** (solid lines) together with simulated results (dashed lines). Devices with 30 min VPP PEDOT:Tos films exhibited colours from blue-green to orange, while a further increase of polymerization time to 60 min pushed the colours further into the red, with a reflectance peak at around 670 nm for the 10 min UV-treated sample. Converting the spectra to CIE chromaticity coordinates indicates good colour tuneability (**Figure 2c**). As discussed above, the purity of the red colours is somewhat limited due to the secondary peak in the blue, although to a lesser extent compared with only tuning the thickness. We also note a limited coverage in the blue region, which is due to interband transitions of Au and corresponding limited reflection of the bottom mirror at short wavelengths. Although Ag or Al mirrors would provide improved reflection at short wavelengths, they led to poor polymer film coverage due to chemical reactions with the precursors (**Figure S4**). To jointly optimize mirror properties and polymer film quality, we therefore replaced the Au mirror with 100 nm Al topped with a 3 nm Cr adhesion layer and only a very thin Au layer (7 nm, see mirror reflection spectra in **Figure S5**). **Figure 2d** presents reflection spectra for UV-treated PEDOT:Tos films prepared on this new Al/Cr/Au mirror, indeed revealing a broad structural colour range that includes blue and violet-blue. The $1^{st}$ order reflectance peak could be tuned from 410 nm to 660 nm, almost covering the whole visible range as also clear from the CIE chromaticity diagram (**Figure 2e**). Numerical simulations for additional polymer thicknesses confirm these findings (**Figure S6** and **S7**). As predicted above, the colour gamut obtained by UV exposure (*i.e.,* combined modulation of both thickness and permittivity was better than expected based on only variations in film thickness (**Figure S8** and **S9**). The devices showed clear but relatively low angle dependence, with only small shifts in reflectance peak positions for incident angles below 40° (**Figure S10** and **S11**). In addition to UV exposure and polymerization times, other processing parameters, including precursor recipe, could also be used to modulate the final structural colours (**Figure S12**). In **Supplementary Note A**, we particularly discuss effects of UV exposure on surface roughness and how an additional annealing step could reduce roughness while reducing tuneability.

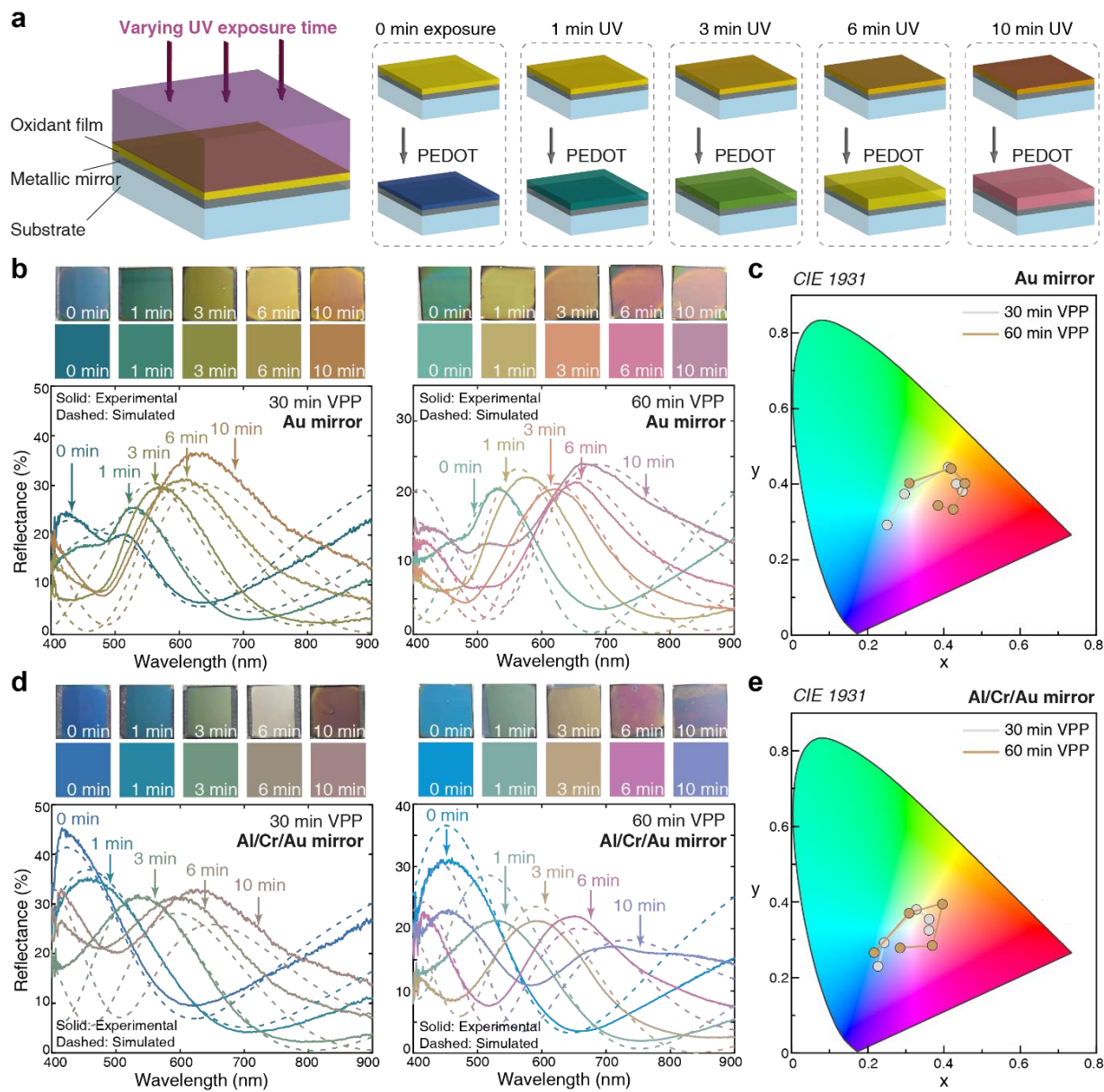

**Figure 2 | Experimental demonstration of UV light controlled structural colours. a,** Fabrication of UV treated PEDOT by different UV light exposure time. **b-e,** Devices made by UV-treated PEDOT films with various exposure time on 100 nm Au mirrors (**b**) and 7 nm Au/3 nm Cr/100 nm Al mirrors (**d**) and their corresponding coordinates distribution in *CIE 1931* chromaticity diagram (**c** and **e**). Five UV exposure times are used: 0, 1, 3, 6, and 10 min. In **b** and **d**, the left column is for 30 min VPP and the right for 60 min VPP; the top panel shows the sample photographs (with size of 2.5 cm by 2.5 cm) in the 1st row and pseudocolours obtained from their CIE coordinates based on experimental reflectance curves (bottom panel) in 2nd row. Solid curves are experimental data while the dashed ones are obtained from numerical simulation.

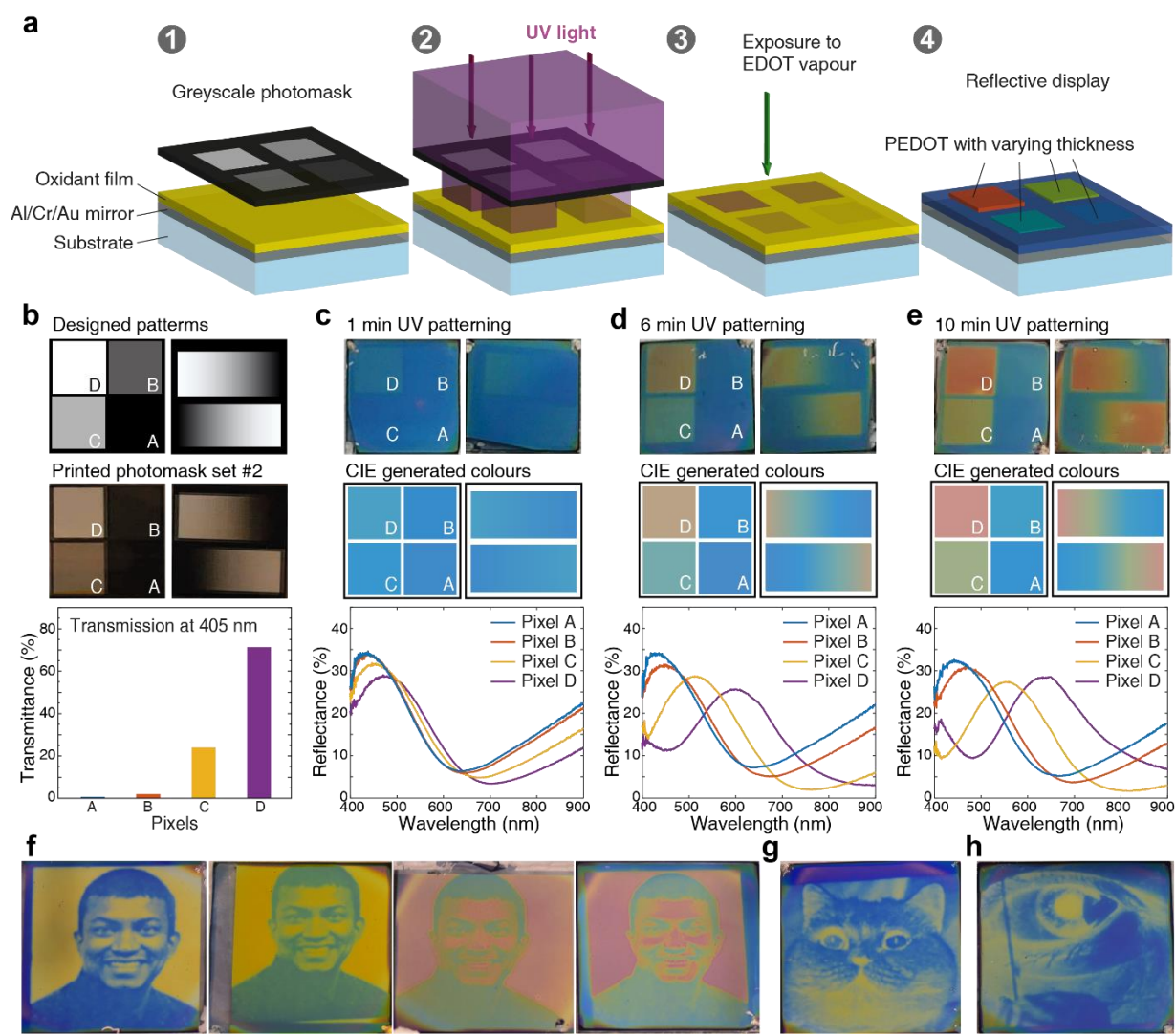

**Figure 3| Reflective structural colour images based on UV-patterned PEDOT on Al/Cr/Au mirror. a**, Schematic process flow of the UV patterning technique. **b**, Designed greyscale patterns with separate pixels and gradient bars (top), photographs of printed photomasks (middle, set #2, set #1 and #3 are shown in **Figure S13**), and corresponding transmittance of pixel A-D at 405 nm (bottom). **c-e**, UV-patterned PEDOT devices using the printed photomask set #2 at different UV exposure times (1, 6, and 10 min). The polymerization time was 30 min. The first, second, and third rows are device images, CIE pseudocolours, and reflectance curves, respectively. The different pixels (A to D) are indicated in both device photographs and CIE generated colour images. **f-h**, UV-patterned PEDOT devices made using grey-scale photomasks of a previous lab member (**f**), a cat (**g**), and an eye (**h**). The different images in **f** correspond to different UV exposure and polymerization times using the same photomask, showing the possibility to tailor the structural coloured images.

The effect of UV treatment on the conducting polymer is determined by the total UV dose[24], which means that modulation of polymer thickness and permittivity can be controlled via the UV light intensity in addition to exposure time. Similar strategies recently enabled fabrication of greyscale electrochromic polymer devices[25,26]. For our structurally coloured devices, insertion of greyscale photomasks during UV exposure should enable the production of multiple colours on the same substrate using just a single exposure step (**Figure 3a**). To verify this, we used greyscale photomasks with both pixels and gradient bar patterns that were printed on transparent plastic films by normal office printers (**Figure 3b**). We used four pixels with transmittance varying from above 70 % to below 1 % at 405 nm, the wavelength of the UV light source. 1 min UV exposure time (30 min VPP on Al/Cr/Au mirrors) led to only moderate variations in colour between the different pixels and across the gradient bars (**Figure 3c**). By contrast, longer exposure times (presented for 6min and 10 min in **Figure 3d** and **3e**, respectively) enabled production of colours from blue to red on the same device (also see **Figure S13**). With a single UV exposure, the concept managed to generate all three primary RGB colours, and the gradient bars further illustrate the possibility for gradual fine-tuning of the structural colours across the visible and at high spatial resolution.

We then utilize greyscale images as photomasks (**Figure 3f-h** and **Figure S14**), which resulted in multi-coloured pictures that precisely reproduce the original photomasks at high resolution. Varying the UV patterning or polymerization time, we were able to further alter the colour map of the same picture. Compared with other approaches to generate structurally coloured images, the one-step UV patterning approach reduces fabrication complexity and offers large scale production with good colour quality, good tuneability and high resolution without need for expensive equipment. There is also plenty of room for studying and further improving the concept by varying different parameters, including the bottom mirror, conducting polymers, wavelengths and intensity of the UV light, and other possible precursor deposition methods (*e.g.,* ink jet and spray printing). The simple structure and facile fabrication process also permit us to form images on flexible substrates. Devices made on poly(ethylene terephthalate) or PET substrates with Au mirrors showed high quality and durability, with no observable damage upon bending (**Figure S25**).

To unveil the mechanisms involved in generating structural colours with conducting polymer thin films on metal mirrors, we analyse the electric field and absorbed power distributions for a device with 170 nm thick 0 min UV-treated PEDOT film on the Al/Cr/Au mirror (**Figure 4a**). At the reflectance maximum, the device generates a constructive interference pattern above the polymer film, while the reflectance minimum shows a much weaker pattern due to destructive interference (**Figure 4b**). These results are associated with lower absorption by the cavity at the reflectance maximum than at the reflectance minimum, which also shows predominant absorption in the top part of the polymer film where the electric field is higher (**Figure 4c**). Similar results were found for devices made of PEDOT:Tos at other thicknesses and treated at different UV exposure times (**Figure S15**), with field and absorption distributions varying with resonance order. Combined with the observed red-shift with increasing thickness and refractive index, the trends are similar for those expected based on constructive interference between light reflected by the air/polymer interface and the polymer/metal interface. However, the air/polymer interface has relatively low reflectance (less than 5% in the visible, see **Figure S16**) and is therefore expected to contribute only moderately to interference effects. To verify this point, we simulated the same device immersed in water to further reduce the refractive index contrast and thereby the reflection at the first (water/polymer) interface. This led to an overall reduction in reflectance features, but otherwise largely remained spectral lineshape, electric field and absorbed power distributions (**Figure 4d-f**). We further replaced water by an artificially created material with identical real refractive index to that of the conducting polymer layer but with zero imaginary part, which is expected to produce no reflection at the top surface of PEDOT:Tos. This system showed essentially identical results as for the device immersed in water, confirming that the structural colour does not require any reflection at the top polymer interface but is instead related to material absorption combined with standing waves formed between incident light and light reflected at the bottom mirror (**Figure S17**). This finding is of high relevance for practical applications, because it allows us to immerse the devices in electrolytes with only little suppression of the structural colours. In turn, electrolyte immersion should enable electrochemical tuning of the energetics of the polymer thin films and corresponding *in situ* control of the optical response of the devices.

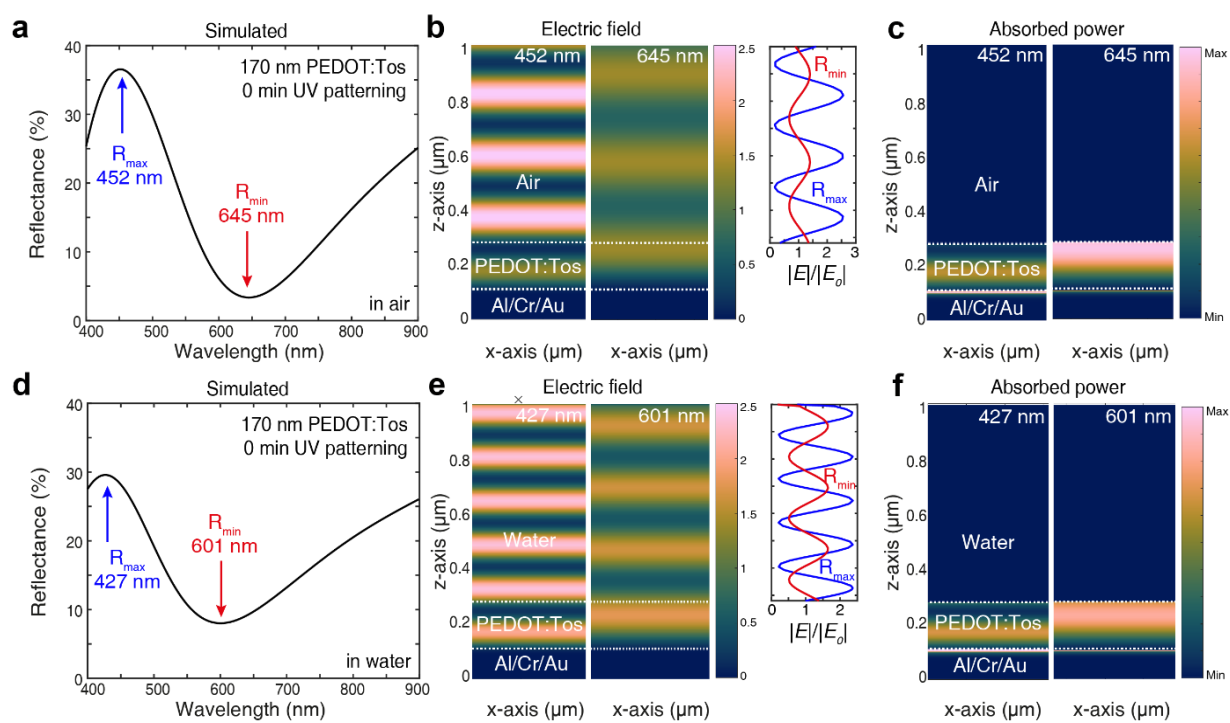

**Figure 4 | Mechanism study of UV-treated PEDOT:Tos devices. a-c**, Device made of 170 nm PEDOT with 0 min UV exposure is used as the model system: reflectance spectra (**a**), electrical field distribution at the reflectance maximum (452 nm) and minimum (645 nm) and the relative electric field strength along *z*-axis above UV-treated PEDOT:Tos (**b**), and absorbed power distribution (**c**). **d-f**, Identical device shown in **a-c** but with water as surrounding media and its reflectance spectra (**d**), electric field (**e**), and absorbed power distribution (**f**). The interfaces of air or water/PEDOT:Tos, and PEDOT:Tos/metallic mirror are indicated by white dashed lines.

For *in-situ* electrochemical switching, we inserted structurally coloured devices (using Au mirrors to ensure electrochemical stability) into a flow cell (**Figure 5a**) filled with electrolyte (1M KCl aq.). We connected the devices to the working electrode (W.E.) and used a Pt electrode as both counter and reference electrode (C.E. and R.E.) in a two-electrode setup. Electrochemical cyclic voltammetry showed rectangular shapes for conducting polymer films (homogenous films, not patterned) treated at different UV exposure times (**Figure 5b** and **Figure S18**). The asymmetry for the 10 min UV sample and pinch-off at negative bias could be due to weakened redox-tunability of samples after prolonged UV exposure[25]. By applying a positive bias to the devices, we could oxidize (or dope) the conducting polymer films, and thus reduce the absorption in the visible (**Figure S19**). This modulation led to increased reflectance and brighter structural colours for both UV-treated and non-treated devices (**Figure 5c-e**). On the contrary, applying a negative bias reduces (or dedopes) the polymer, which results in enhanced absorption in visible and reduced absorption in the near infrared. Such negative bias therefore dimmed the devices, which presented dark red or brown colours (**Figure 5c** and **Figure S18**). Devices using Al/Cr/Au mirrors showed improved colour tunability, with colour changing from green to purple, and finally to red, although with the disadvantage of deteriorating in the electrochemical environment (**Figure S20**). Regardless of mirror-type, the change of reflectance and device colour occurred within seconds. When removing the bias, the samples slowly returned to their original state, with variations for different UV exposure times (or electrical/ionic conductivity). In addition to the electrochemically induced reflectance modulation (as demonstrated in **Figure 4d**), we observed a red-shift of colour when the environment of the devices changed from air to water, especially for samples with longer UV treatment times. This effect is likely caused by swelling of PEDOT:Tos[23], which increases the film thickness and alters the resonance condition for the device. Despite the reduced reflectance and colour shifts, the polymer's ability for *in situ* switching shows potential for practical applications like multi-colour labels, displays, and adaptive camouflages. We demonstrate the possibility to switch structurally coloured images using a UV-patterned PEDOT:Tos device showing a portrait of one of the authors (**Figure 5f**, immersed in a beaker instead of the flow cell). By tuning the electrochemical bias, we could reversibly control the emergence and disappearance of the picture, showing dark violet to green and yellow colours. Further improvements and practical devices may be aided by introducing solid-state electrolytes.

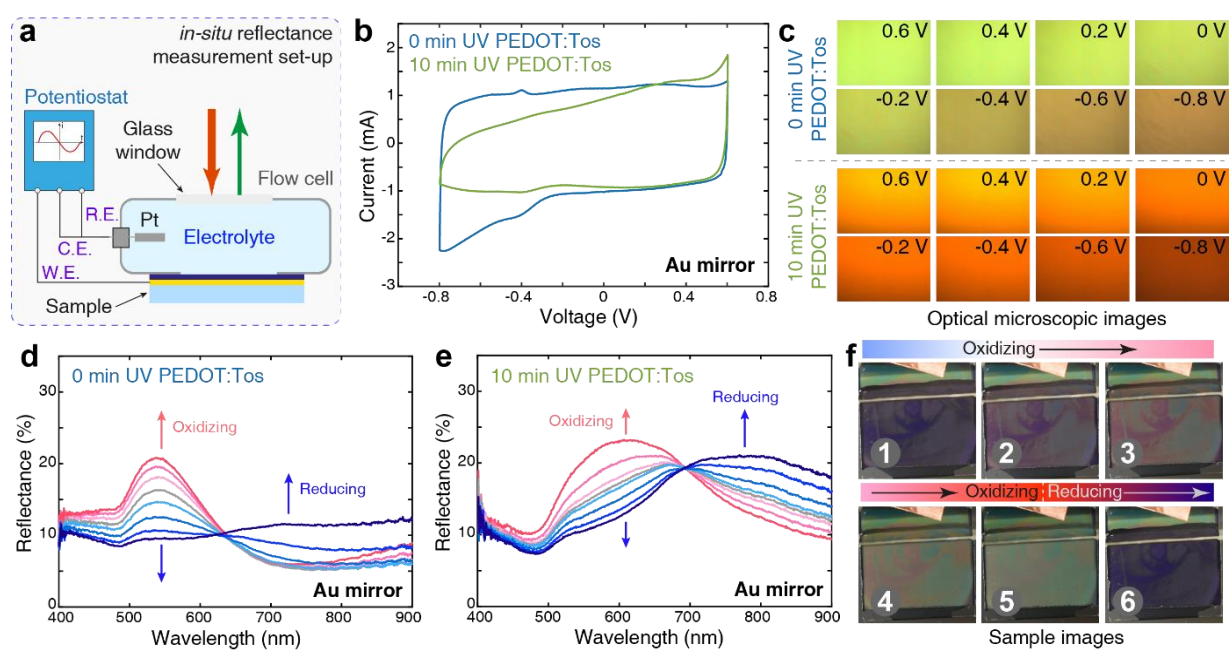

**Figure 5| Electrochemical responses of UV-patterned PEDOT displays. a**, Schematic of the flow cell used for electrochemical measurement. R.E. is reference electrode, C.E. is counter electrode, and W.E. is working electrode. **b**, Cyclic voltammetry of UV patterned PEDOT:Tos films on 100 nm Au mirrors (0 min and 10 min UV exposure). The scan rate was 100 mV/s. **c**, Optical microscope images of UV-patterned PEDOT devices at different electrochemical bias (top: 0 min UV-treated PEDOT, and bottom: 10 min UV-treated PEDOT). **d** and **e**, *in-situ* electrochemical tuning of the reflectance of 0 min UV-treated PEDOT (**d**) and 10 min UV-treated PEDOT (**e**) devices. Blue curves indicate negative bias (reducing the polymer) and red curves indicate positive bias (oxidizing the polymer). The intensity of red and blue colours represents the increase in positive and negative electrochemical bias indicated by the arrows. Grey colour curves indicate the zero electrochemical bias. **f**, Electrochemical tuning of a device with a 10 min UV-patterned image (VPP time of 30 min).

In this study, we demonstrate structural colours using thin conducting polymer films on metal mirrors and multi-colour display devices fabricated *via* VPP combined with a single UV patterning step. The concept provides wide colour gamut owing to UV-controlled modification of both polymer permittivity and thickness, which jointly modulates the structural coloration. Taking advantage of the electrochemically tuneable properties of the conducting polymer, the concept also offers reversibly switchable colour images. We believe that the presented concept can aid the development of next-generation backlight-free energy-efficient displays. Future research may optimise UV patterning parameters and explore different polymers, mirrors and cavity structures, to further improve colour purity and electrochemical switching properties.

## Methods

**Thin film deposition and device fabrication.** PEDOT:Tos thin films were deposited *via* vapour phase polymerization (VPP) in a vacuum chamber. Detailed information about the process and setup can be found in our previous studies[21,23]. Briefly, substrates coated with precursor [or oxidant, consisting 2 g of Clevios™ C-B 54 V3 (Heraeus, Germany), 2 g of tri-block co-polymer PEG-PPG-PEG (Sigma-Aldrich), and 5 g of water-free ethanol (99.7 wt%)] were introduced into a vacuum chamber filled with EDOT monomer vapour (by heating EDOT droplets at 60 °C). After certain polymerization time (30 or 60 min), post-baking process at 70 °C for 2 min, and ethanol/DI water washing and nitrogen flow drying steps, PEDOT:Tos films can form on top of the substrates (glass or glass coated with metallic mirrors). For UV light exposure technique, the precursor (oxidant) films were placed under a UV lamp at wavelength of 405 nm and power of 20 mW/cm$^2$ for different times. Greyscale photomasks or blank quartz photomask can be used during the exposure to create patterns or pure colours in the finally obtained PEDOT films. The greyscale photomasks used in the study were printed on transparent plastic films by normal office inkjet printers (RICOH, Japan) for simplicity. For display devices, glass coated with metallic mirrors (100 nm Au or 7 nm Au/3 nm Cr/100 nm Al deposited by thermal evaporation) were used as substrate and the size is normally 2.5 cm by 2.5 cm. Free-standing PEDOT film transfer for glass substrates coated with Ag can be done with PEDOT:Tos films deposited on silicon wafers and washed in a DI water bath and more details can be found in our previous studies[23]. To deposit UV-exposure-free PEDOT:Tos films with different thicknesses, we reduced spin-coating speed of precursor films or polymerization time to prepare thin films (below 150 nm) and increased polymerization time or used layer-by-layer method to make thick films (above 200 nm).

**Reflectance spectra and optical microscope.** The reflectance spectra of the devices were recorded by a home-made micro-spectroscopic setup[27]. Briefly, samples were illuminated by a 50 W halogen lamp attached to an optical microscope (Nikon Eclipse L200 N, Japan) and the reflected light was split into two parts at the camera port, one for optical microscope image capturing and the other for spectra *via* an optical fiber (with a diameter of 100 µm) connected to a spectrometer (QePRO, Ocean Optics, USA). For all measurements, an objective lens of 10x magnification was used. Incidence angle dependence reflectance and scattered reflectance spectra were recorded by a home-made setup shown in **Figure S13** and **S15**, where a set of concentric wheels were used to precisely control the angles with resolution of 10°. The devices were placed at the rotation center of the inner wheel and the detector arm was mounted at the edge of outer wheel, while collimated white light was incident on the devices at fixed angles. Rotating both wheels, incident angles or scattered light angles can be tuned. Reflected light was collected by an optical light guide connected to a spectrometer (Andor Shamrock 303i with a Newton CCD detector). Device images were taken either by the optical microscope or mobile phone (Huawei Mate 20).

**Optical, electrical, and structural characterization.** Optical extinction spectra were measured using a UV-vis-NIR spectrometer PerkinElmer Lambda 900 in the range 2500 nm to 400 nm at the step of 1 nm. UV-vis-NIR ellipsometry were performed using a J. A. Woollam Co. RC2® SE with 4-5 incident angles. All measurements were carried out in normal ambient at room temperature. UV-treated PEDOT thin films were deposited on double-side polished sapphire substrates (Prime Wafers, Netherlands) and for each sample, at least two spots were measured. The details of measurements and data analysis can be found in our previous studies[21]. Sheet resistance of the films, $R_s$, was measured *via* four-point probe method using a Signatone Pro4 S-302 resistivity stand and a Keithley 2400 SourceMeter. Film thickness $t$ were measured by a surface profiler (Dektak 3st Veeco) and confirmed by the thickness determined from ellipsometry. The electrical conductivity can be calculated by the formula *σ = 1/($R_s$ × t)*. Atomic Force Microscope (AFM) height and phase images were collected using a Veeco Dimension 3100 in tapping mode.

**Optical numerical simulations.** Numerical simulations of the reflectance spectra, the electric field, and the absorbed power distribution were carried out *via* the finite-difference time-domain method using the commercial software Lumerical FDTD solutions. Optical refractive index of UV-treated PEDOT were extracted from the ellipsometry measurements. UV-treated PEDOT film on top of metallic mirrors and substrates were illuminated by a plane wave light source at normal incidence. Antisymmetric and symmetric boundaries were used for the x-y axis (in-plane direction of PEDOT films) while perfectly matched layers were used for the z axis (out-of-plane direction of PEDOT films). Mesh size used in the simulation was typically $1 \times 1 \times 1$ nm$^3$. Stackrt command in Lumerical Script based on analytic transfer matrix method for multi-layer stack was also used to verify the numerical simulation results. Results for artificial material of PEDOT:Tos with zero imaginary part was simulated using an equivalent setup using COMSOL-Multiphysics.

**Electrochemical characterization.** *in-situ* electrochemical reflectance spectra were collected via a flow cell illustrated in **Figure 5a**. In this two-electrode set-up, the device was connected to the working electrode of an electrochemical potentiostat, while a Pt electrode was used as both reference/ counter electrode. The electrolyte used in the study is 1M KCl water solution. For cyclic voltammetry, the scan rate was 100 mV/s. For the reflectance measurement, the whole flow cell was placed into the home-made micro-spectroscopic set-up (there is a glass window on the front side of the flow cell for light transmission). For each electrochemical bias, chronocoulometry was used and the spectra were collected after about 30 seconds for stabilization. *in-situ* electrochemical absorption spectra were taken in an identical way in a quartz cuvette placed inside the sample holder of the UV-vis-NIR PerkinElmer spectrometer in the range 900 nm to 400 nm at the step of 2 nm.

## Acknowledgements

The authors acknowledge financial support from the Swedish Foundation for Strategic Research (SSF), the Knut and Alice Wallenberg Foundation, the Swedish Research Council (VR), the Wenner-Gren Foundations, and the Swedish Government Strategic Research Area in Materials Science on Functional Materials at Linköping University (Faculty Grant SFO-Mat-LiU No 2009 00971). M.P.J. is a Wallenberg Academy Fellow.

## Author contributions

M.P.J. and S.C. conceived and designed the project. S.C. and G.C. prepared the film. S.C. fabricated the device, and performed the optical, electrical and structural characterizations with the assistance of J.E. S.R. and S.C. performed the electrochemical measurements of the devices. R.S., S.C., S.G. performed the micro-spectroscopic measurements. S.R. and S.C carried out optical numerical simulations. P.K., V.S., V.D., and S.C. performed ellipsometry characterizations. S.C. and M.P.J. wrote the manuscript. All co-authors reviewed and commented on the manuscript.

## Competing interests

The authors declare no competing interests.

**Supplementary Information for**

# Redox-tunable structural colour images based on UV-patterned conducting polymers

*Shangzhi Chen[1], Stefano Rossi[1], Ravi Shanker[1], Giancarlo Cincotti[1], Sampath Gamage[1], Philipp Kühne,[2,3], Vallery Stanishev[2], Isak Engquist[1], Magnus Berggren[1], Jesper Edberg[4], Vanya Darakchieva[2,3] and Magnus P. Jonsson[1]\**

[1] Laboratory of Organic Electronics, Department of Science and Technology (ITN), Linköping University, SE-601 74 Norrköping, Sweden
[2] Terahertz Materials Analysis Center (THeMAC), Department of Physics, Chemistry and Biology (IFM), Linköping University, SE-581 83 Linköping, Sweden
[3] Center for III-Nitride Technology, C3NiT-Janzèn, Department of Physics, Chemistry and Biology (IFM), Linköping University, SE-581 83 Linköping, Sweden
[4] RISE Research Institutes of Sweden, Bio- and Organic Electronics, Bredgatan 35, SE-602 21 Norrköping, Sweden
\* Correspondence: magnus.jonsson@liu.se

# Part I: Supplementary Notes

## Supplementary Note A

We observed an increase in surface roughness (from 13 nm to 47 nm) with increasing UV exposure time (**Figure S21**), although normal incidence reflectance spectra show a rather small percentage of non-specular scattered light (less than 1 percent in the visible, see **Figure S22**). The somewhat rough 10 min UV-patterned PEDOT films contained both small dots in size of tens of nanometers and concave or convex surface in sizes of 5-10 µm, which start to emerge at 3 min UV exposure as observed by optical microscopes (**Figure S23**)[1,2]. Similar features have been reported by Murphy *et al.*[3] and considered as a result of humidity-induced crystallization of iron salts (Fe(Tos)$_3$) in the oxidant film. Crystallization of iron salts weakens the coordination effect between iron ions and the tri-block co-polymer additives (see **Methods**) and therefore creates more nucleation sites for film growth, largely modulating the polymerization kinetics[4]. In our case, the UV exposure step was carried out at atmospheric conditions without any special inert gas protection and samples could be heavily affected by the humidity, especially for samples with long UV exposure times. Optical microscopy reveals two different regions: areas showing the macroscopically observed colours and small areas with dark colours that sometimes generate additional reflectance peaks or widens the reflectance peak width corresponding reduction in colour purity (**Figure S24**).

To avoid this crystallization effect, we thermally annealed samples after UV patterning but prior to VPP, which largely supressed the formation of iron salt crystallites (see 'baked oxidant' in **Figure S23** and **S24**). However, the devices made using this strategy showed more limited colour tuneability (**Figure S25**), where the reflectance peak only shifted between 450 nm (blue) and 570 nm (yellow-green). Further characterization of these films reveals that the thermal annealing step largely diminished the patterning effect by reducing the thickness variations and partly also the variations in the optical extinction (**Figure S26**). Ellipsometry data (**Figure S27** and **28**) show lower variations in both real and imaginary refractive index upon UV exposure of the baked oxidants, further corroborating the findings.

Therefore, there is currently a trade-off between colour purity (reflectance peak width) and colour gamut, which may be further optimized *via* processing parameters, such as oxidant recipes, polymerization conditions (*e.g.,* substrate temperature), and precise local environment humidity control.

# Part II: Supplementary Figures

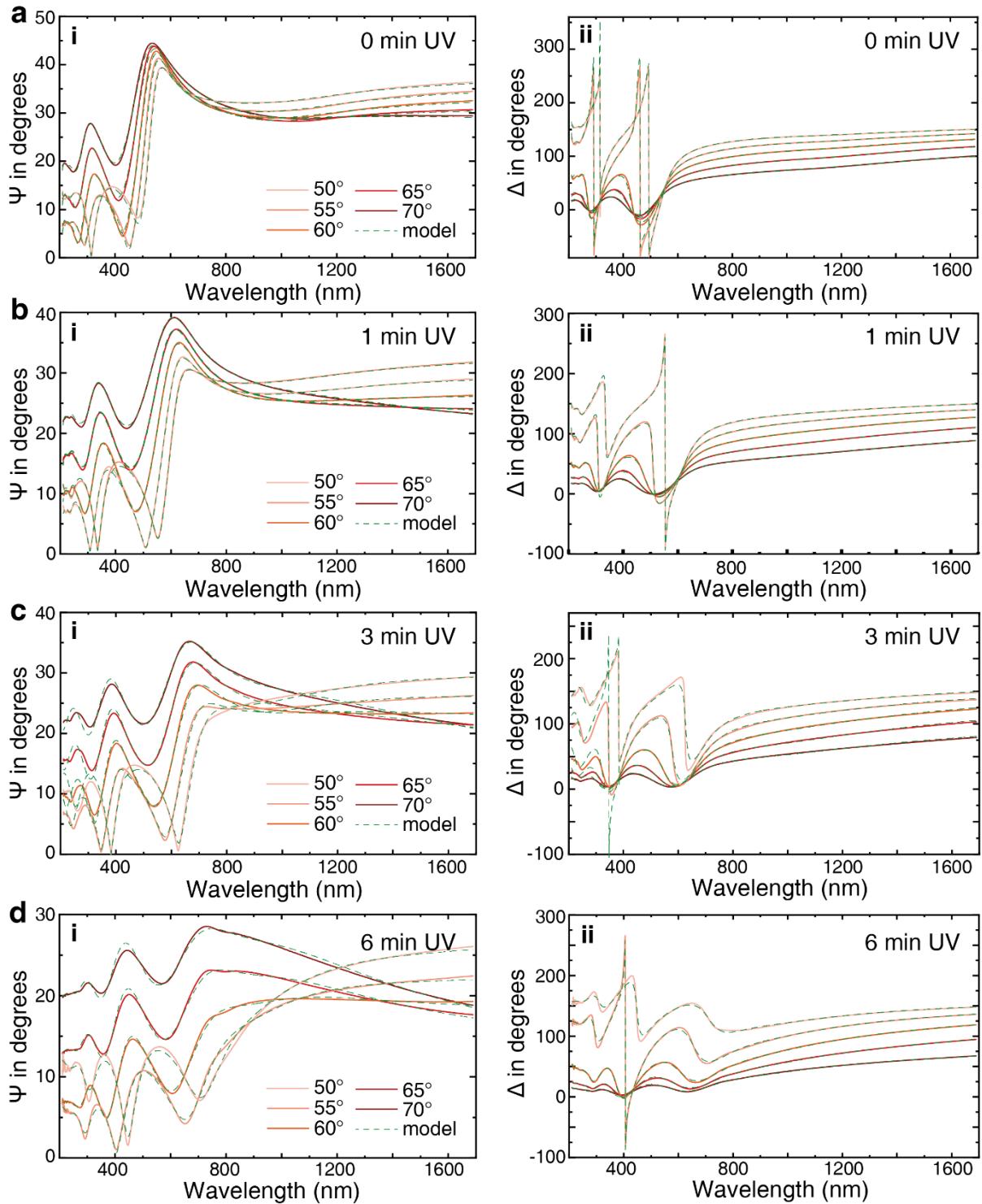

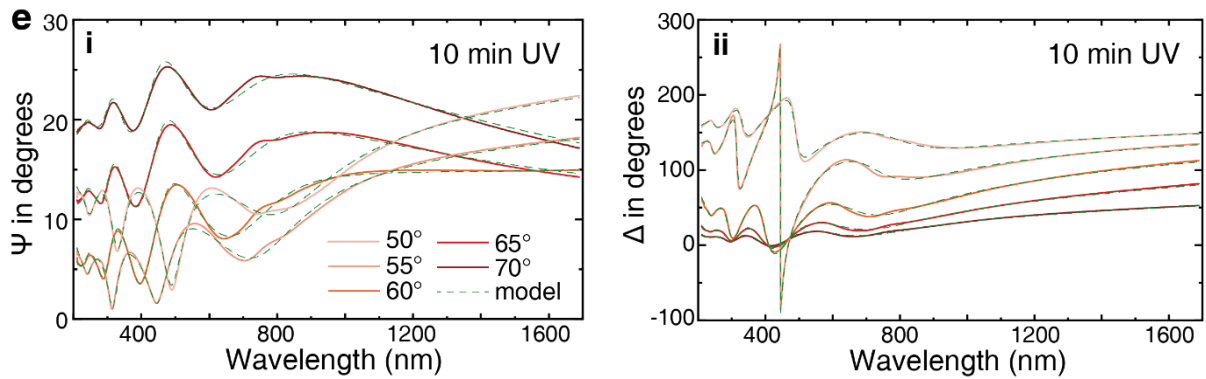

**Figure S1 | Spectroscopic ellipsometry spectra for UV-treated PEDOT thin films.** Five UV exposure times are used: 0 min (**a**), 1 min (**b**), 3 min (**c**), 6 min (**d**), and 10 min (**e**). For each measurement, five incident angles are used: 50°, 55°, 60°, 65°, and 70°. Drude-Lorentz model[5] is used to fit for the data. The experimental measured data (**i** for $\psi$ and **ii** for $\Delta$) and model fits are plotted in red solid and green dashed curves, respectively. The UV-treated PEDOT films are deposited on sapphire substrates. The complex refractive index dispersions derived from the Drude-Lorentz model are shown in **Figure 1c, 1d,** and **S2**.

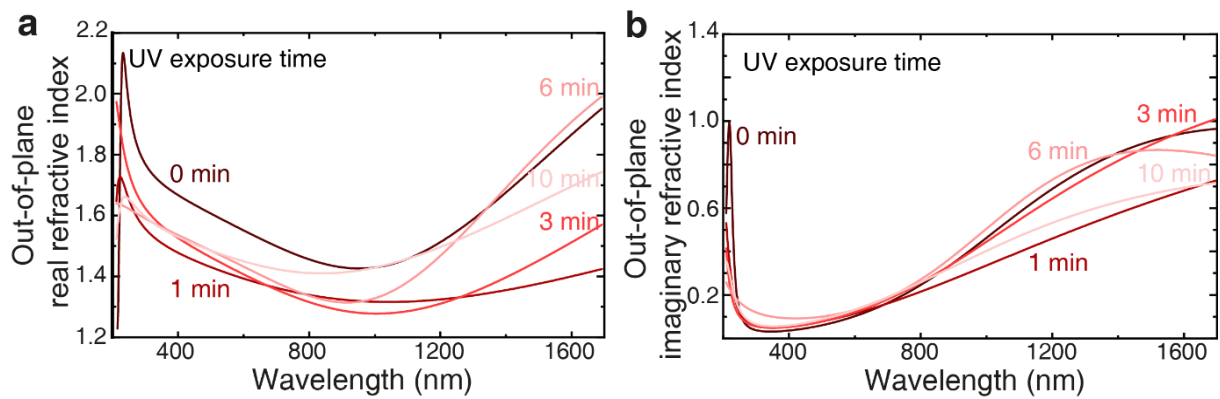

**Figure S2 | Out-of-plane refractive index of UV-treated PEDOT thin films. a**, Real refractive index. **b**, Imaginary refractive index. Their in-plane counterparts are shown in **Figure 1c** and **1d**. UV exposure time used are 0, 1, 3, 6, and 10 min.

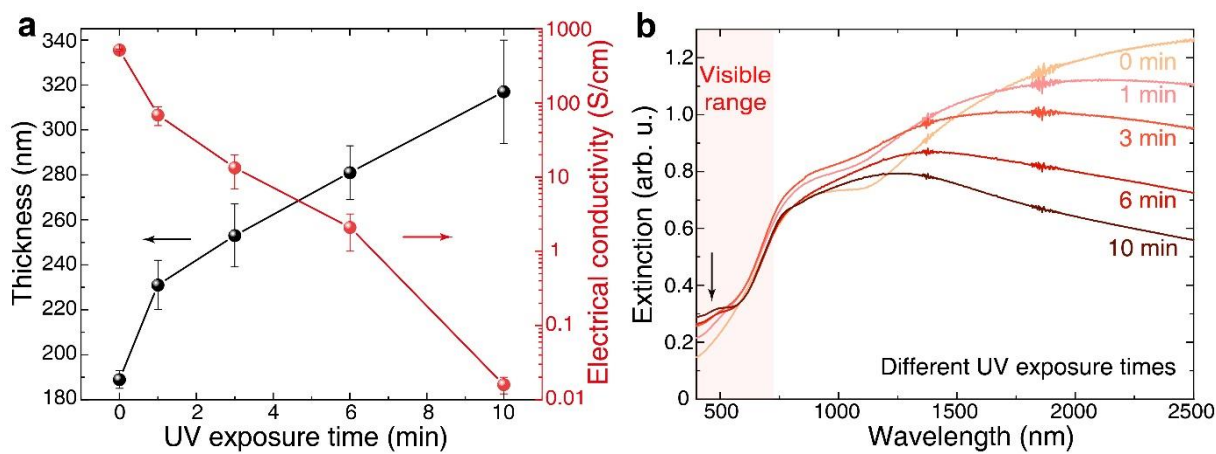

**Figure S3 | Properties of UV-treated PEDOT thin films. a,** Thin film thickness and electrical conductivity with error bars, **b**, Extinction spectra of PEDOT thin films prepared by exposure time ranging from 0 to 10 min and polymerization time of 60 min. The samples are prepared using microscope glass slides as substrates and the finally obtained thickness values can vary depending on the type of substrates. The light red shaded area in **b** represents visible range. The arrow in **b** indicates the additional extinction shoulder due to the existence of short-chain polymers[1,2]. With the increase of UV exposure time, the extinction in near infrared region is decreasing and short-chain polymer signal intensity at short wavelengths is increasing.

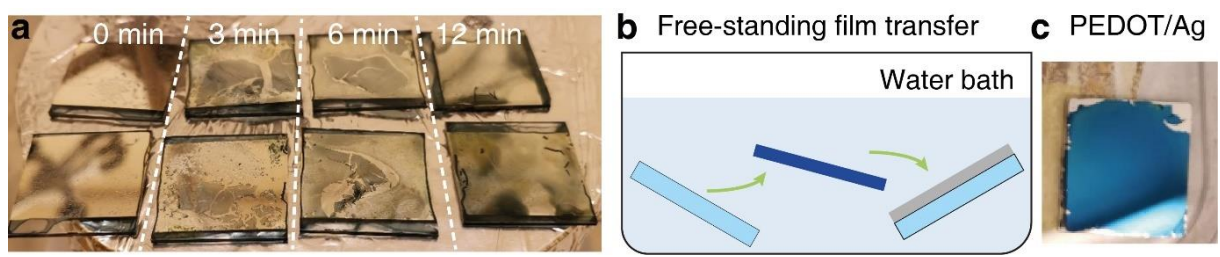

**Figure S4 | Devices using silver bottom mirrors. a**, UV-treated PEDOT films deposited on Ag bottom mirrors. Four different UV exposure intervals are used: 0, 3, 6, and 12 min. It is found that oxidant can directly react with the Ag bottom mirror upon its deposition and no uniform film can be polymerized in all samples. In addition, silver film is hydrophobic and no oxygen-plasma treatment should be used to improve hydrophilicity due to the potential oxidation of silver layer. **b**, Free-standing film transfer strategy is used to re-deposit as-formed PEDOT films to new substrates (see details in **Methods** section)[6]. PEDOT film can be delaminated in water bath and thus oxidant can avoid direct contact with silver film. Pristine PEDOT film on silver layer is shown in **c**. However, for UV-treated PEDOT samples, the films show strong adhesion to the original substrates and can hardly be transferred to the new silver substrates. Therefore, this method cannot be used for preparing UV-treated PEDOT on silver bottom mirrors.

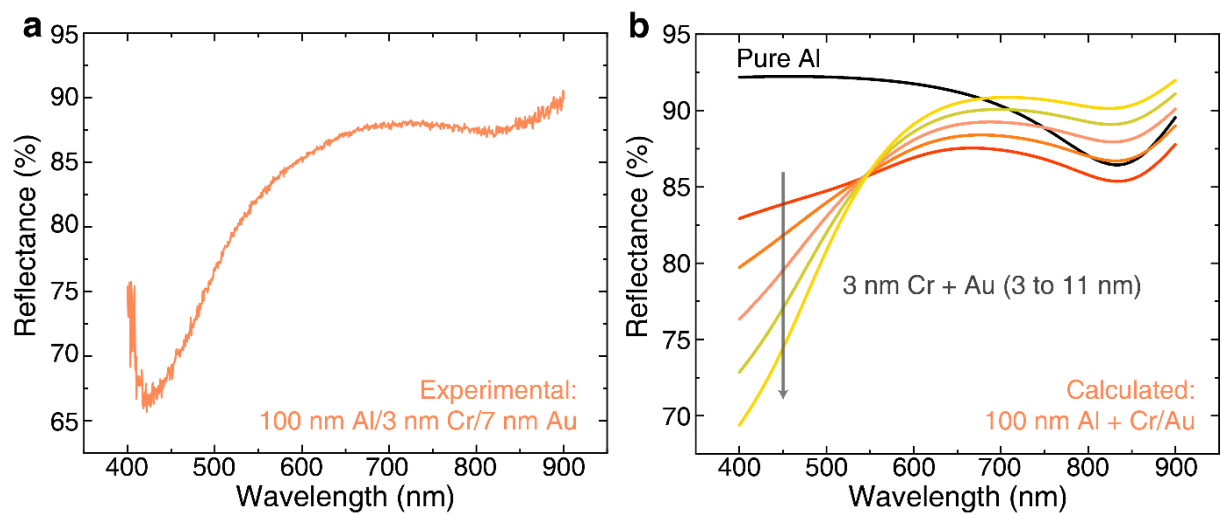

**Figure S5 | Reflectance spectra of different metallic mirrors. a**, Experimental reflectance curves of 7 nm Au/3 nm Cr/100 nm Al mirror. The mirror was deposited on glass substrate. **b**, Simulated reflectance curves of 100 nm Al with different thickness combination of Cr and Au layers. The thickness of Au is ranging from 3 nm to 11 nm with a step of 2 nm as indicated by the grey arrow. Pure Al curve is plotted in black line. The overall shape of the curve from calculation matches well with experimental curve. The difference between experiment and calculation may be due to surface roughness and poor coverage of the thin Cr and Au layer. Also, the optical parameters of ultrathin metal layer might deviate from that of the bulk.

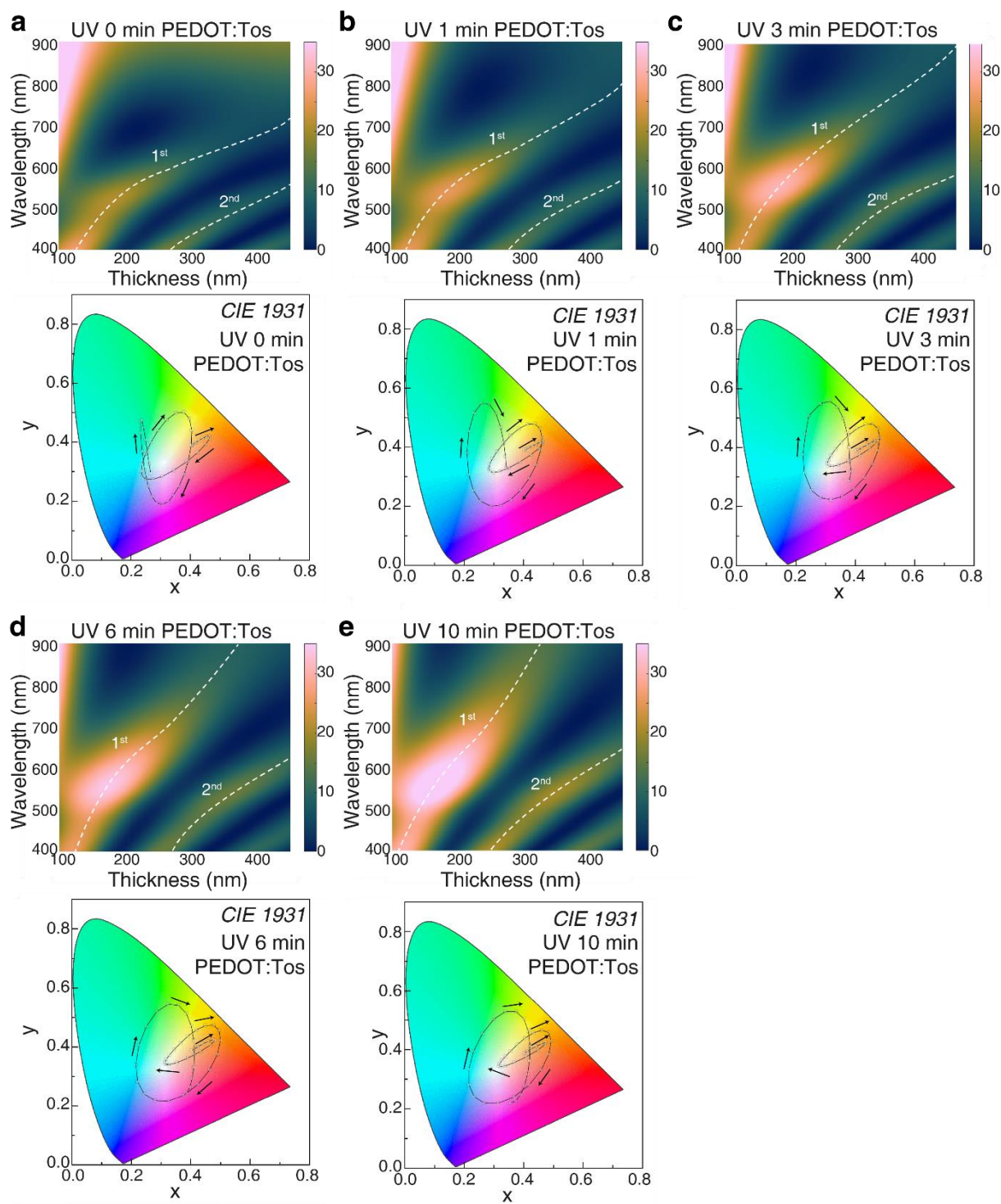

**Figure S6| Simulated reflectance of UV-treated PEDOT:Tos on 100 nm gold bottom mirror.** PEDOT:Tos thin films with different UV exposure times are used: 0 min (**a**), 1 min (**b**), 3 min (**c**), 6 min (**d**), and 10 min (**e**). The top panel is 2D heat map of simulated reflectance with respect to film thickness and the bottom panel is the distribution of CIE coordinates for these devices in *CIE 1931 xy* chromaticity diagram.

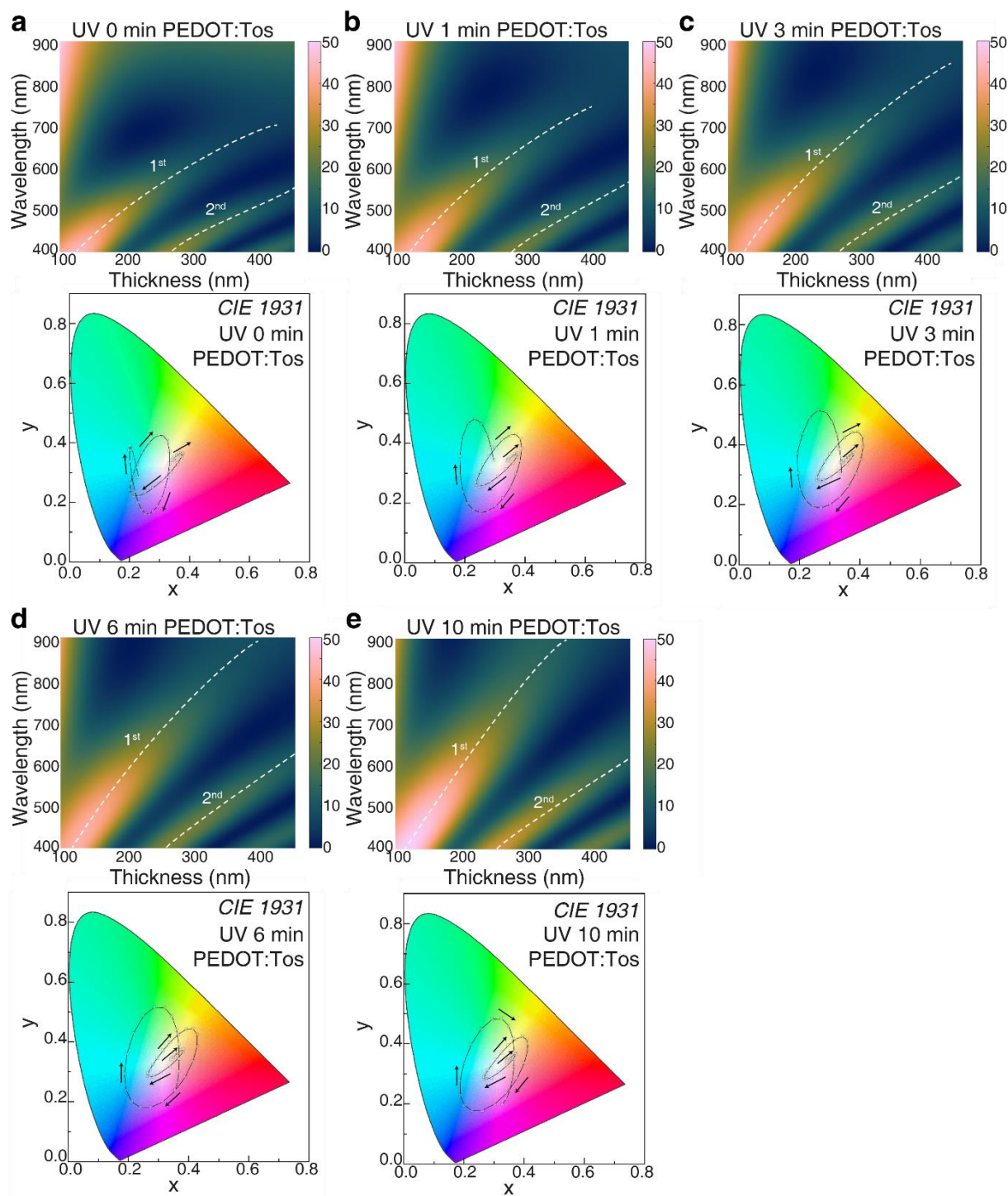

**Figure S7| Simulated reflectance of UV-treated PEDOT:Tos on 7 nm Au/3 nm Cr/100 nm Al bottom mirror.** PEDOT:Tos thin films with different UV exposure times are used: 0 min (**a**), 1 min (**b**), 3 min (**c**), 6 min (**d**), and 10 min (**e**). The top panel is 2D heat map of simulated reflectance with respect to film thickness and the bottom panel is the distribution of CIE coordinates for these devices in *CIE 1931 xy* chromaticity diagram.

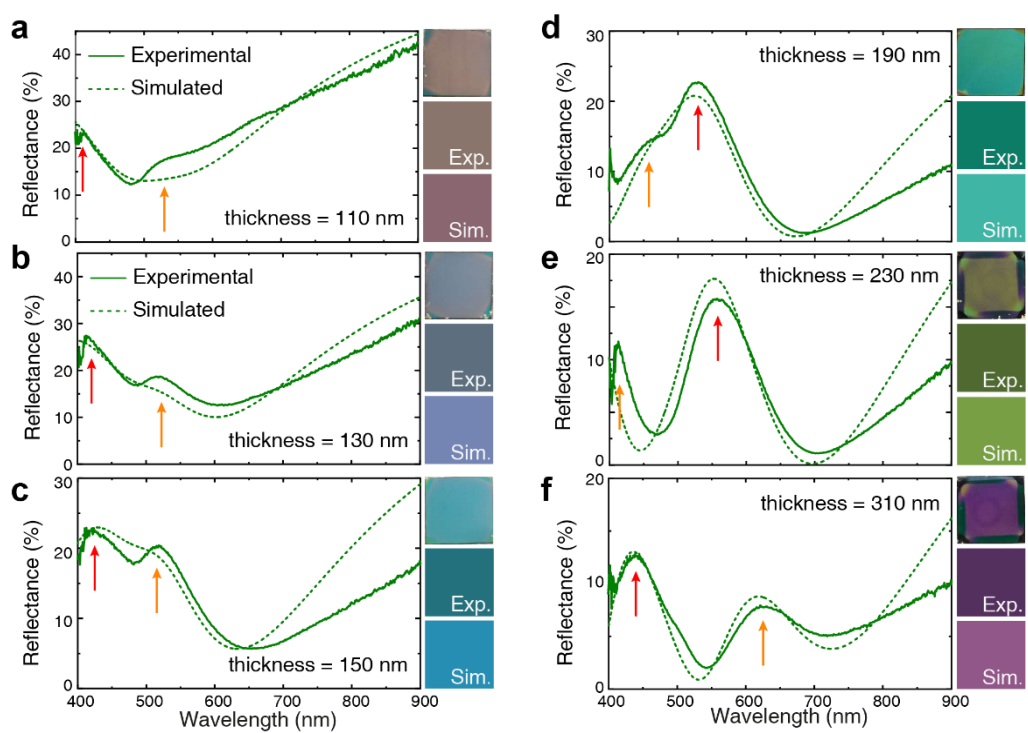

**Figure S8 | Structural colours of PEDOT by only tuning film thicknesses.** PEDOT:Tos with no UV treatments were prepared with different thicknesses: 110 nm (a), 130 nm (b), 150 nm (c), 190 nm (d), 230 nm (e), and 310 nm (f). 100 nm bottom Au mirrors were used. Experimental (solid) and simulated (dashed) reflectance spectra are presented. The panels to the right of each graph display sample photographs (1st row) and pseudocolours obtained from CIE coordinates based on experimental (2nd row) and simulated reflectance curves (3rd row). The generated reflective colours from pink-grey (thickness of 110 nm) to yellow-green (thickness of 230 nm), with corresponding primary reflectance peak positions varying from 410 nm to 570 nm, agreeing well with simulated results. A 2nd order reflectance peak becomes dominant for larger thicknesses of about 250 nm (with the 1st order peak located at around 590 nm), which largely limits pure colour generation at longer wavelengths (*e.g.* yellow and red).

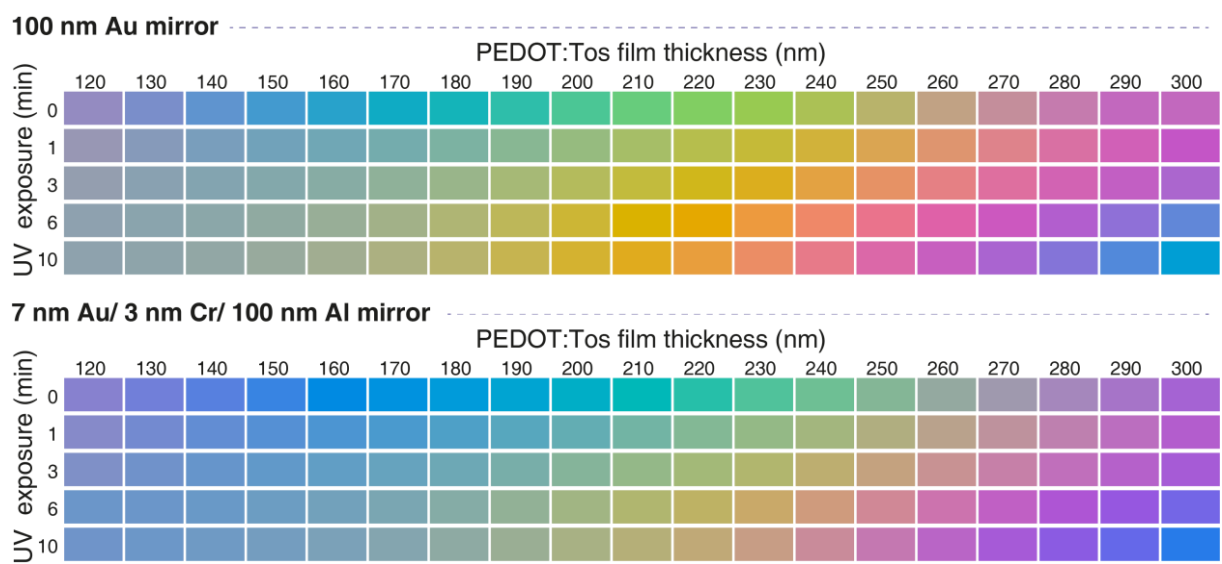

**Figure S9 | Pseudocolours based on the simulated reflectance spectra.** The CIE coordinates are obtained from simulated reflectance spectra (**Figure S6 and S7**) to create pseudocolours. The top panel is for 100 nm Au bottom mirror while the bottom panel is for 7 nm Au/3 nm Cr/100 nm Al bottom mirror. The *x*-axis is the thickness of PEDOT:Tos film and *y*-axis is the UV exposure time. It is obvious that UV exposure (modulating both film thickness and absorption profiles) can widen the colour ranges by creating more longer wavelength colours (*e.g.* yellow and red) than only tuning thickness.

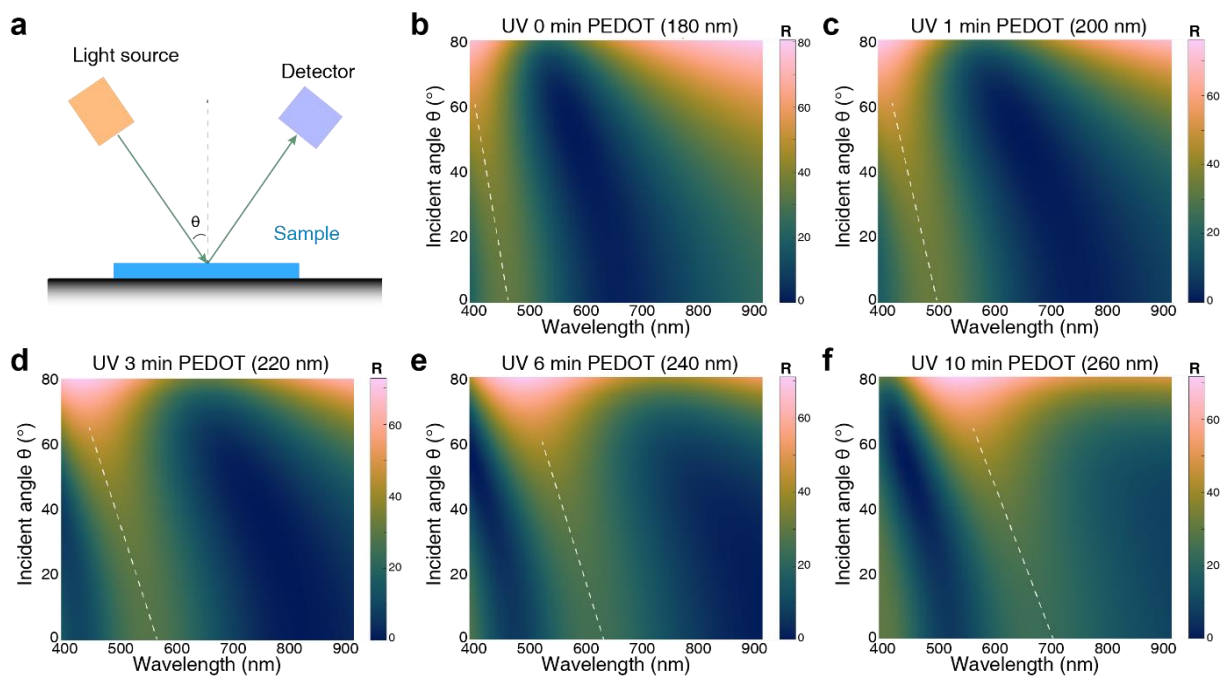

**Figure S10 | The incident angle dependence of calculated reflectance. a,** Illustration of setup analogues to that used for the calculations, where the incident angle θ is defined the angle between incoming (outgoing) ray and surface normal. **b-e**, 2D heat map of reflectance spectra of UV 0 min (**b**), 1 min (**c**), 3 min (**d**), 6 min (**e**), and 10 min (**f**) PEDOT:Tos with respect to the incident angle. It can be observed that the slopes of primary reflectance peak slightly tilt with the increase of incident angle. The reflectance curves were calculated by *stackrt* function of Lumerical FDTD solutions.

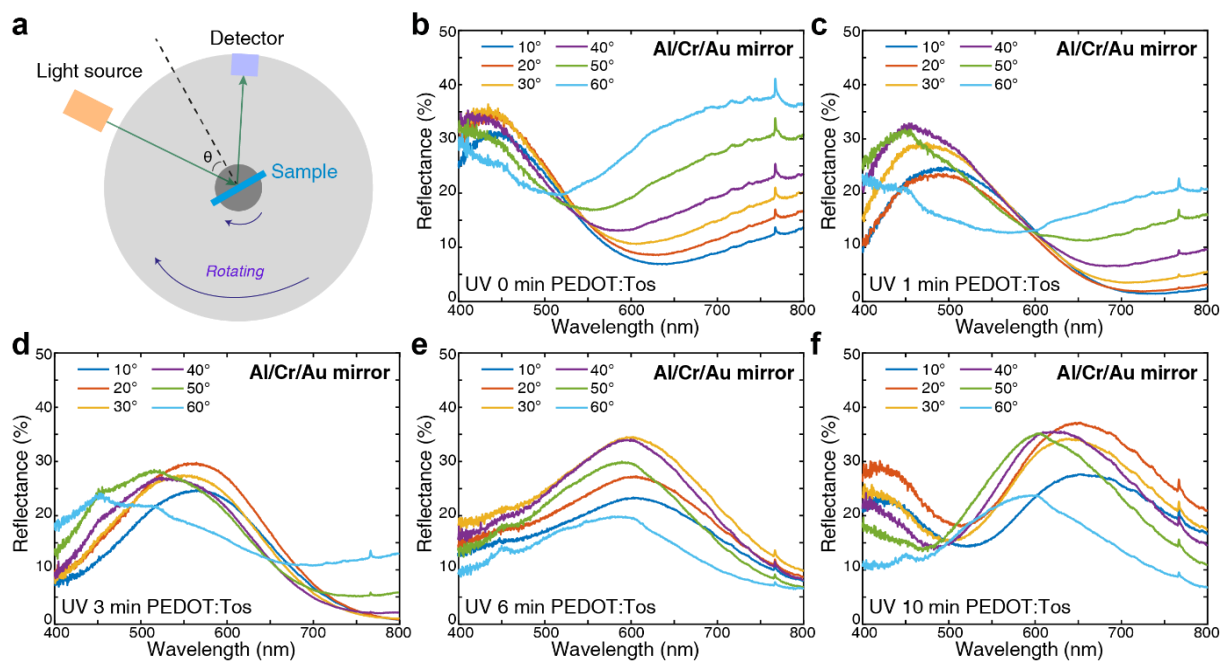

**Figure S11 | The incident angle dependence of experimental reflectance spectra. a,** The set-up of experimental measurements. A set of concentric wheels are used: the inner wheel is used as a sample holder and the outer one is used to place detector. b-f, The incident angle θ dependence of reflectance spectra for 0, 1, 3, 6, and 10 min UV-treated PEDOT:Tos devices.

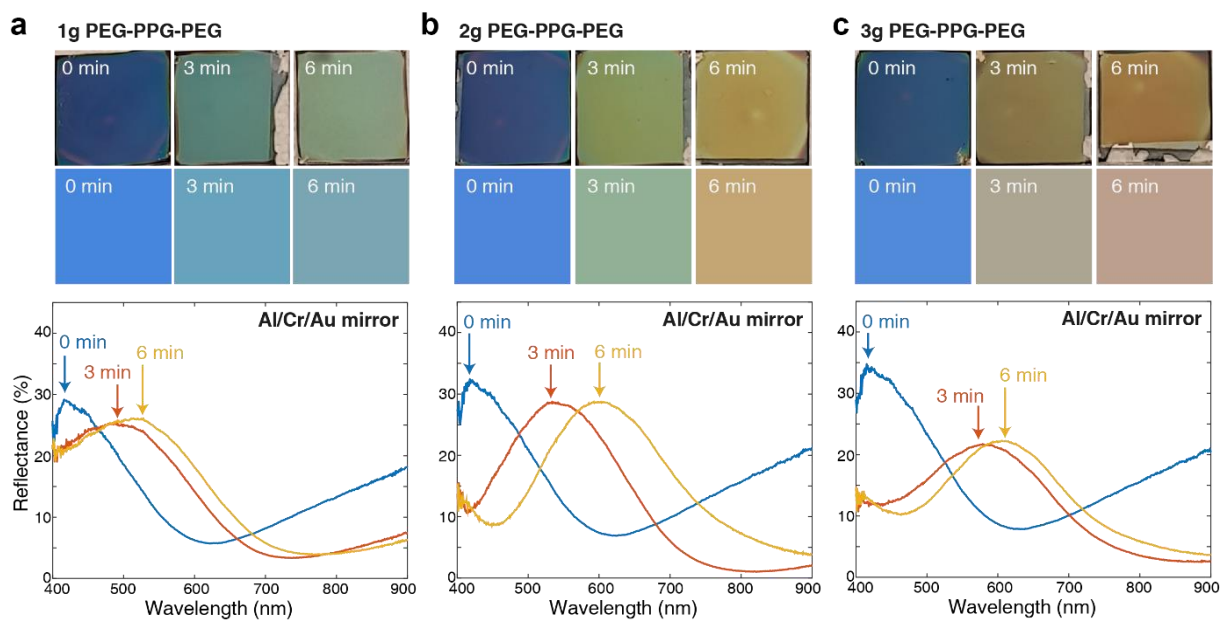

**Figure S12 | Influence of precursor recipes on the reflective colours.** The concentration of tri-block-co-polymer, PEG-PPG-PEG, in the oxidant solution is tuned with different weight. 1g, 2g, and 3g PEG-PPG-PEG, together with 2g of CB-54 and 5g of ethanol are used. a, b, and c show sample photos (1st row), CIE generated colours (2nd row), and corresponding reflectance spectra (3rd row). All samples are based on 7 nm Au/3 nm Cr/100 nm Al bottom mirror. It is clear that by optimizing the concentration of PEG-PPG-PEG, the UV patterning effect can be tuned including changes in both reflectance peak wavelengths and widths.

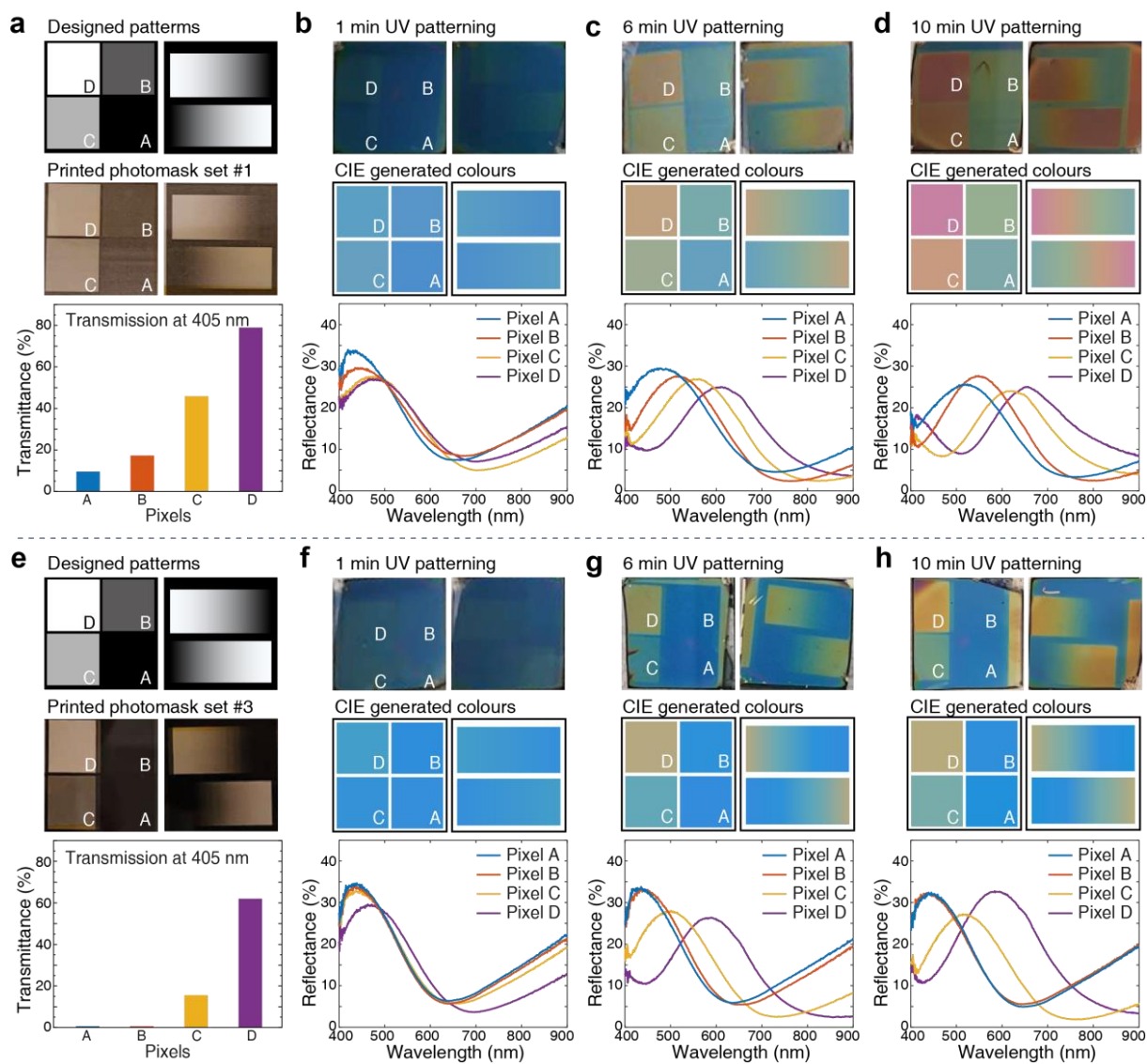

**Figure S13| Greyscale photomasks and the corresponding UV-patterned PEDOT displays. a**, Designed greyscale photomask patterns with pixels and gradient bars (1st row) and its printed version #1 (2nd row). The 3rd row is the transmittance curves for four pixels in the UV-vis. **b-d**, Sample images, CIE generated colours, and corresponding reflectance spectra for 1, 6, and 10 min UV patterned samples. **e-h**, Photomask set #3 and its resulted samples with 1, 6, and 10 min UV patterning. Since photomasks are printed with a normal office printer and it is difficult to achieve all the greyscale colours within a single printing step, we therefore print identical patterns on the same photomasks and attempt to overlap them in order to make pixel C and D darker. We denote the 1-time printed, 2-time printed, and 3-time printed photomasks as set #1, #2, and #3 indicated in **Figure 3b**, **Figure S13a**, and **S13e**.

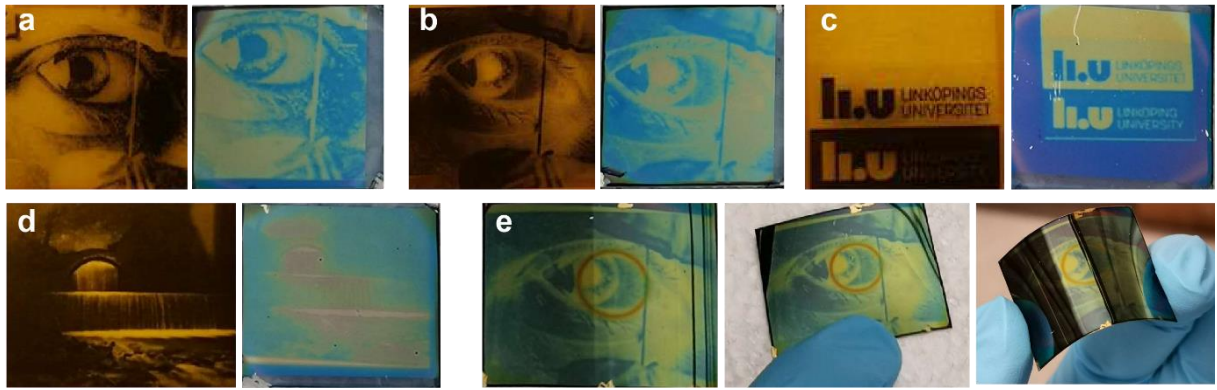

**Figure S14 | UV-patterned PEDOT devices using greyscale photomasks. a-d,** UV patterned PEDOT devices made on glass substrates (left: greyscale photomasks and right: devices). **e**, UV patterned PEDOT devices made on PET substrates (left: device image, middle: device in an original sate, and right: device in a bent state). The red ring is an artifact related to spin-coating of the oxidant on the flexible PET substrate.

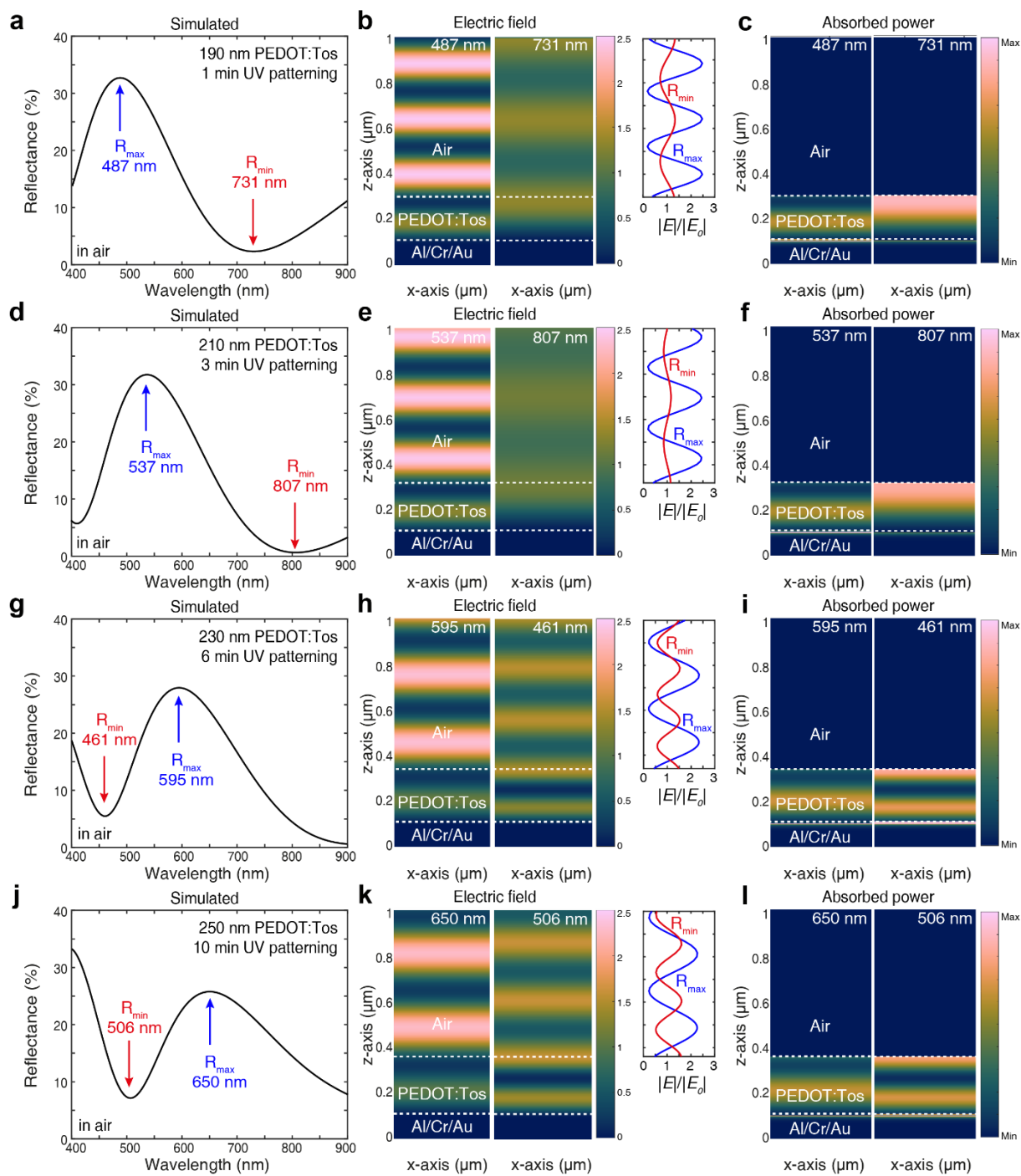

**Figure S15 | Electric field and absorbed power distribution of UV-treated PEDOT devices. a, d, g,** and **j**, Reflectance spectra of 1 min UV patterned (190 nm), 3 min UV patterned (210 nm), 6 min UV patterned (230 nm), and 10 min UV patterned (250 nm) PEDOT devices. **b**, **e**, **h**, and **k**, Electric field distribution at the reflectance maximum (left column), reflectance minimum (middle column), and the relative electric field strength along *z* axis above PEDOT surfaces (right column). **c**, **f**, **i**, and **l**, Absorbed power distribution at the reflectance maximum (left column) and reflectance minimum (right column). Wavelengths for reflectance maximum and minimum of each device are indicated in the reflectance spectra.

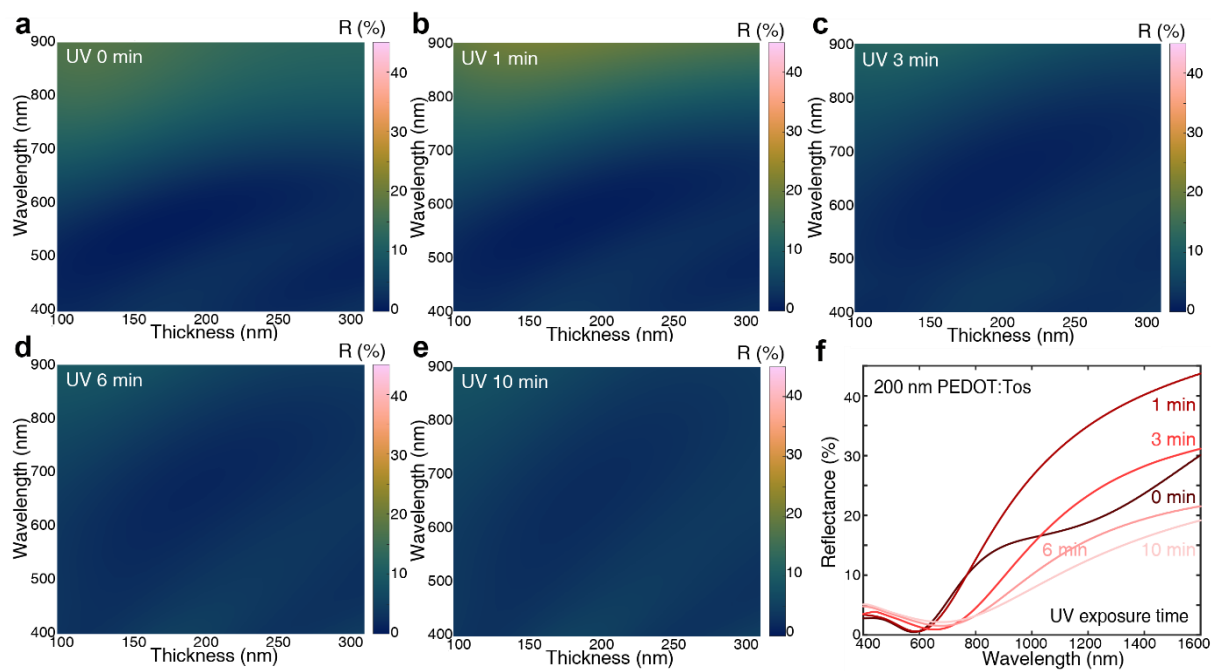

**Figure S16 | Simulated reflectance of UV-treated PEDOT thin films. b-f**, Reflectance 2D heat maps for PEDOT thin films with different thickness for UV exposure time of 0 min (**a**), 1 min (**b**), 3 min (**c**), 6 min (**d**), and 10 min (**e**). **f**, Reflectance curves of 200 nm PEDOT films with different UV exposure time (0, 1, 3, 6, and 10 min) for comparison. The reflectance at short wavelengths has a slight increase with the increase of UV patterning time. In the simulation, PEDOT thin film is placed on top of a glass substrate without any metal bottom mirrors.

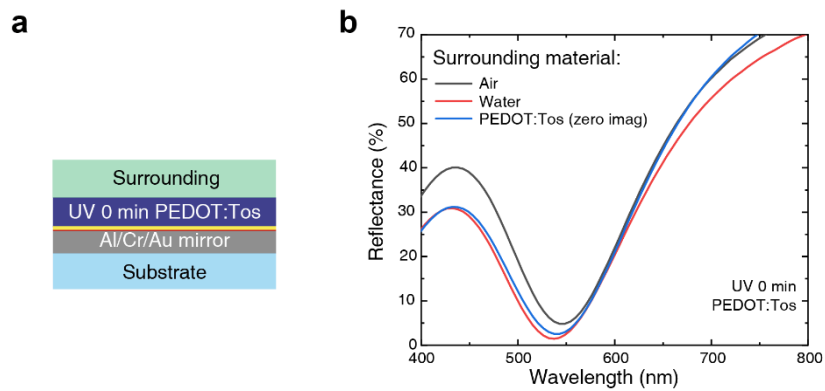

**Figure S17 | The influence of surrounding medium on the device performance. a,** The device structure used for simulation where we change the surrounding materials. **b**, The simulated reflectance spectra using COMSOL. Three different surrounding media are used: air (n = 1), water (n = 1.33), and PEDOT:Tos (zero imag), where the real refractive index of PEDOT:Tos remains but the imaginary part is zeroed.

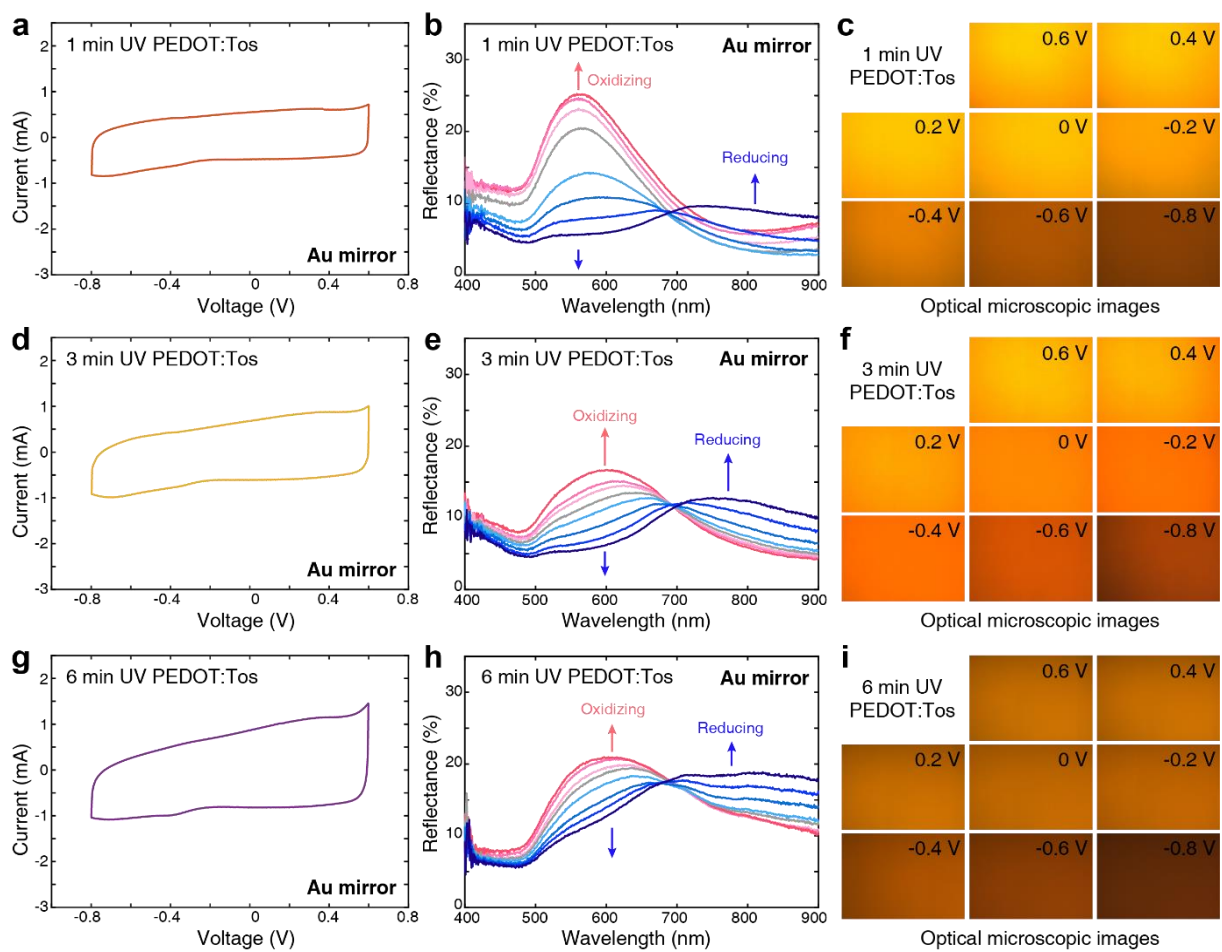

**Figure S18 | Electrochemical responses of UV-treated PEDOT displays. a, d,** and **g,** Cyclic voltammetry of UV patterned PEDOT films on Au mirrors (**a,** 1 min, **d,** 3 min, and **g,** 6 min). The scan rate used is 100 mV/s. **b, e,** and **h,** *in-situ* electrochemical reflectance spectra (**b,** 1 min, **e,** 3 min, and **h,** 6 min). **c, f,** and **i,** Optical microscope images of samples at different electrochemical bias (**c,** 1 min, **f,** 3 min, and **i,** 6 min). The intensity of curve colours represents the absolute values of electrochemical bias (dark red to light red: 0.6 V, 0.4 V, 0.2 V, and dark blue to light blue: -0.8 V, -0.6 V, -0.4 V, -0.2 V). Grey colour curve is for zero electrochemical bias.

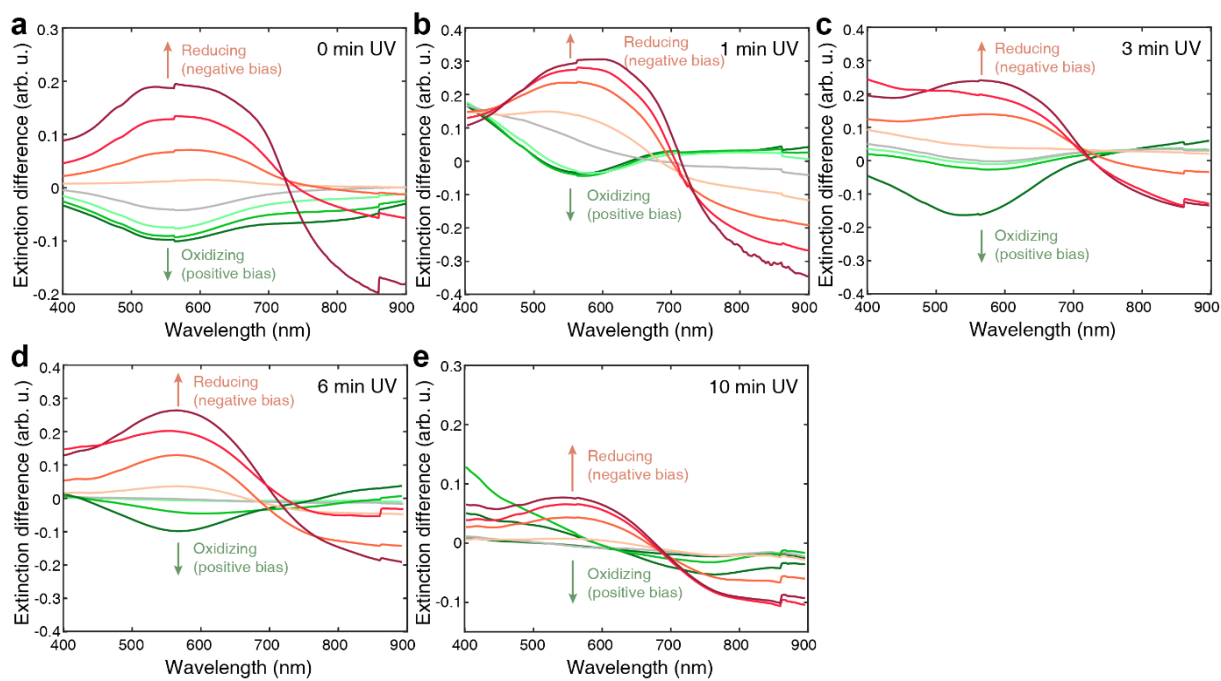

**Figure S19 | *In-situ* electrochemical extinction difference spectra of UV-treated PEDOT:Tos films. a,** 0 min, **b,** 1 min, **c,** 3 min, **d,** 6 min, and **e,** 10 min UV-treated PEDOT:Tos films. The spectra is using the pristine state PEDOT:Tos without electrochemical bias in the electrolyte as the reference, and measuring the net extinction changes at different electrochemical bias. The sum of optical extinction and transmission is 1. Green curves (dark to light: 0.6 V, 0.4 V, and 0.2 V) indicate positive electrochemical bias (oxidizing the polymer) and red curves (dark to light: -0.8 V, -0.6 V, -0.4 V, and -0.2V) indicate negative electrochemical bias (reducing the polymer). All the films are deposited on ITO glasses.

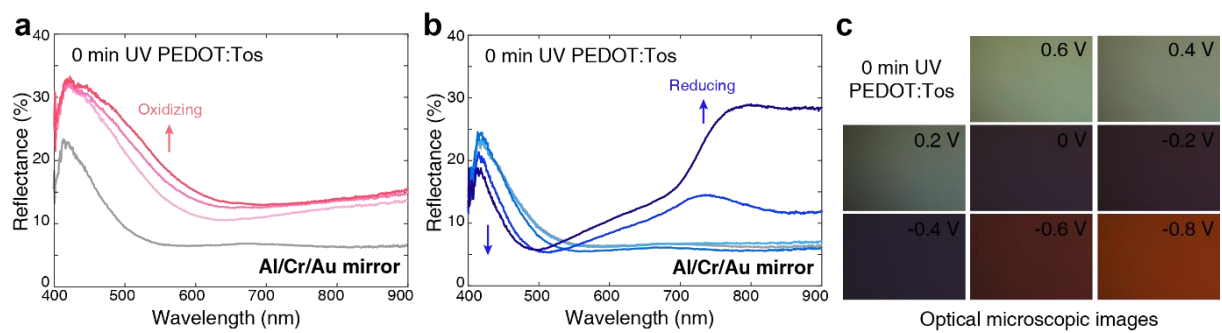

**Figure S20 | Electrochemical responses of UV-treated PEDOT displays on Al/Cr/Au mirror.** **a** and **b,** *in-situ* electrochemical reflectance spectra (**a,** oxidation, **b,** reduction). **c,** Optical microscope images of samples at different electrochemical bias. The intensity of curve colours represents the absolute values of electrochemical bias (dark red to light red: 0.6 V, 0.4 V, 0.2 V, and dark blue to light blue: -0.8 V, -0.6 V, -0.4 V, -0.2 V). Grey colour curve is for zero electrochemical bias. However, the Al/Cr/Au mirror is not stable in the electrochemical environment and starts to fall off from the substrate after several cycles.

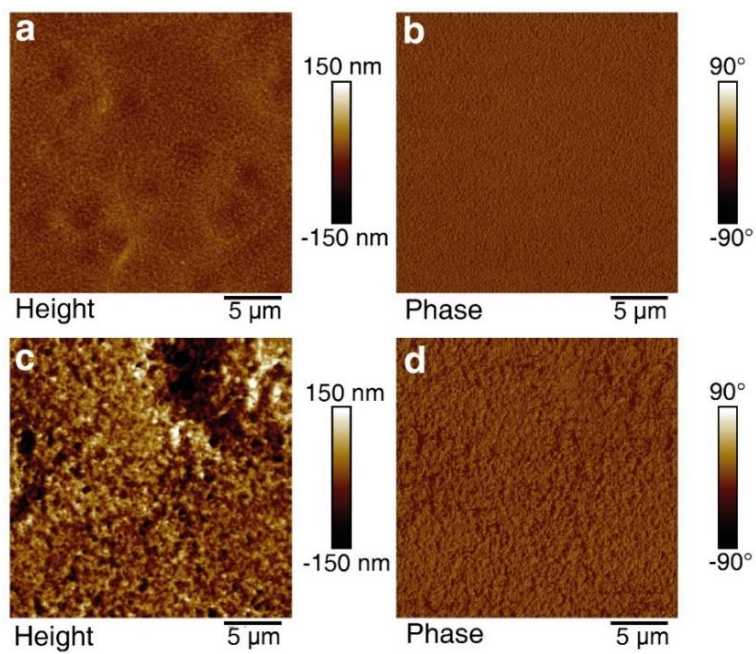

**Figure S21 | AFM images of PEDOT:Tos thin films. a** and **b**, height and phase images of PEDOT thin films prepared by oxidant without UV exposure. The surface roughness is about 12.5 nm. **c** and **d**, height and phase images of PEDOT prepared with oxidant exposed to UV for 10 min. The surface roughness is about 46.9 nm.

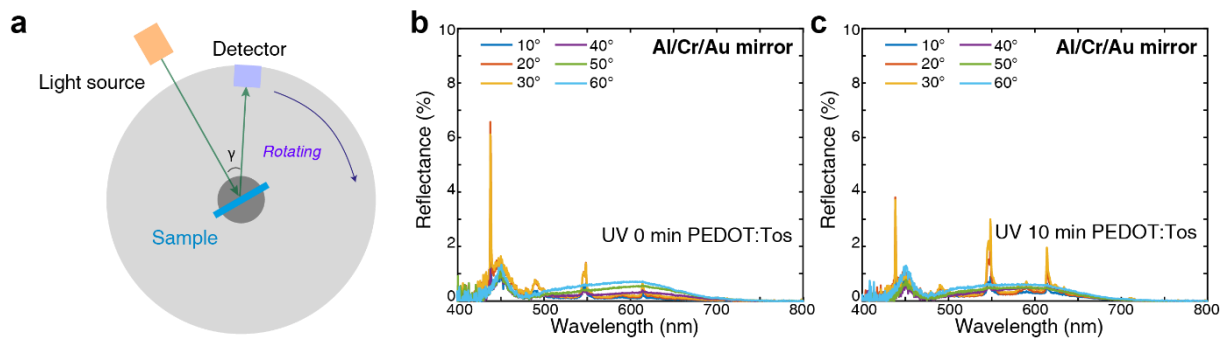

**Figure S22 | Experimental scattered reflectance spectra of UV-treated PEDOT:Tos devices.**
**a**, The set-up of experimental measurements for normal incidence. **b-c**, Reflectance spectra for 0 min (**b**) and 10 min (**c**) UV-treated PEDOT devices. The scattered light was collected at 10° to 60°. The reflectance is mostly below 1 % in the visible.

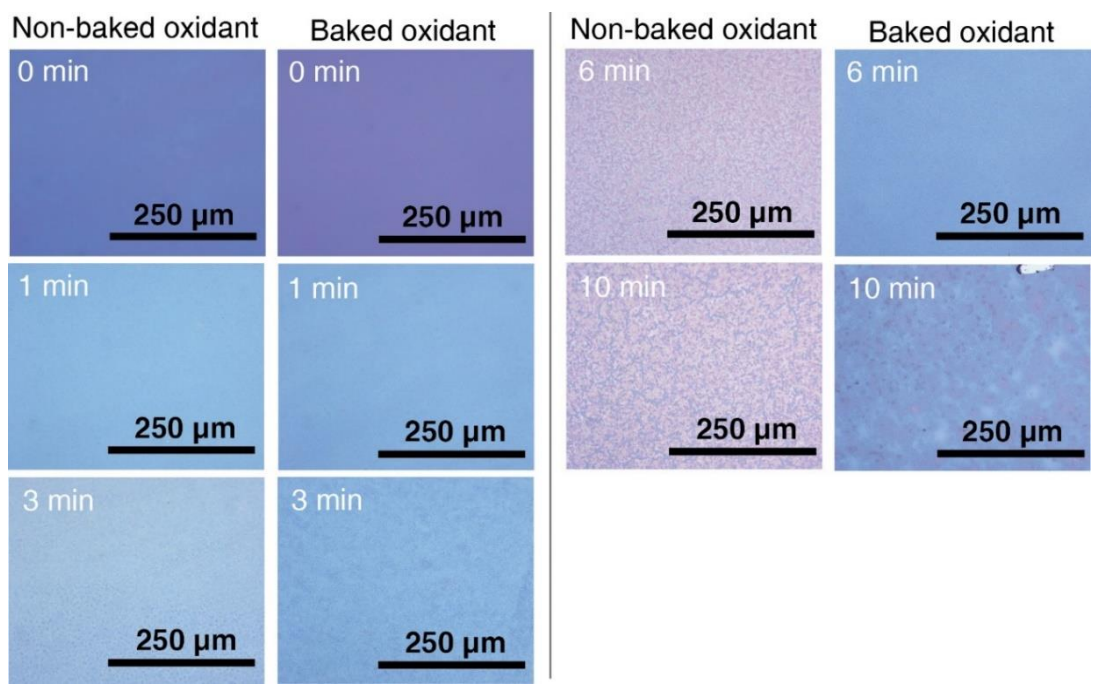

**Figure S23 | Optical microscope images of UV-treated PEDOT:Tos films.** The UV exposure time used are 0, 1, 3, 6, 10 min, respectively. The images were captured in transmission mode. For non-baked oxidant sample, particle-like features start to emerge after 3 min UV exposure while for baked oxidant sample these features are largely suppressed.

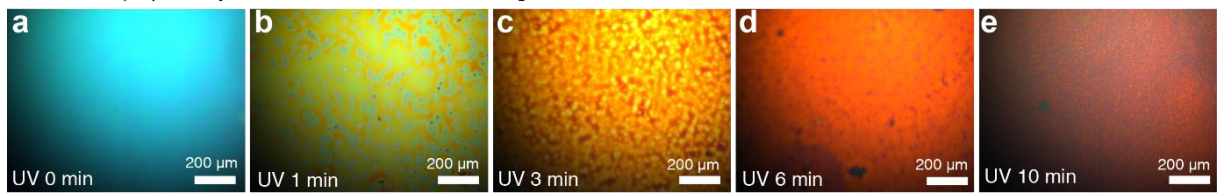
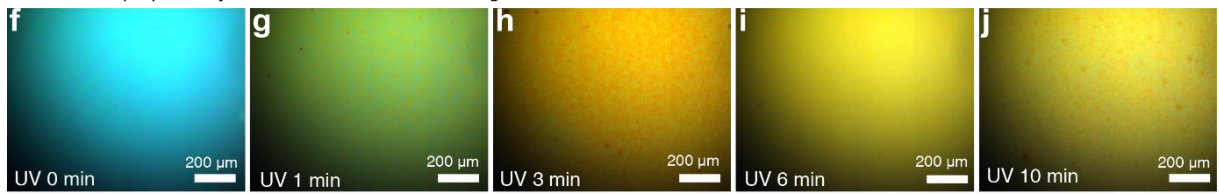

**Figure S24 | Optical microscope images of UV-treated PEDOT:Tos devices. a-e,** UV patterned PEDOT devices made with non-annealed oxidant. **f-j,** UV patterned PEDOT device made with annealed oxidant. The devices were deposited 7 nm Au/3 nm Cr/100 nm Al mirror. UV exposure times used were: 0 min (**a, f**), 1 min (**b, g**), 3 min (**c, h**), 6 min (**d, i**), and 10 min (**e, j**). The microscope images are taken in reflection mode.

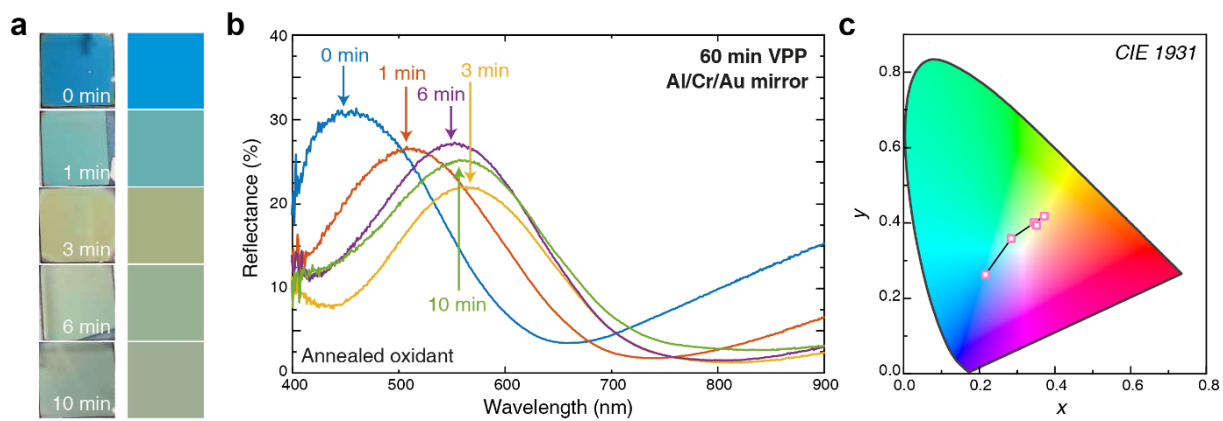

**Figure S25 | UV-treated PEDOT:Tos devices based on thermally annealed oxidant. a,** Device images (left) and pseudocolours from CIE coordinates based on experimental reflectance curves. **b,** Experimental reflectance curves of the devices on 7 nm Au/3 nm Cr/100 nm Au mirror. **c,** The distribution of corresponding CIE coordinates in *CIE 1931* chromaticity diagram. UV exposure times used were: 0 min, 1 min, 3 min, 6 min, and 10 min.

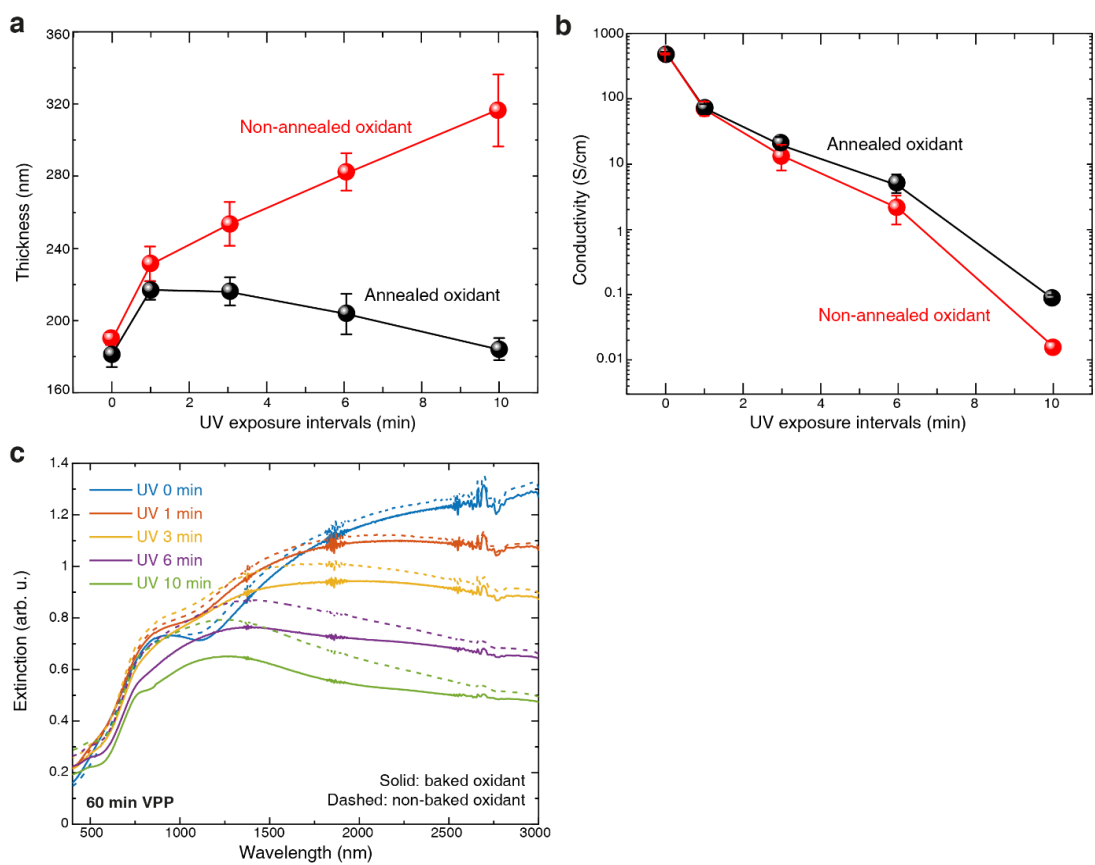

**Figure S26 | Properties of UV-treated PEDOT thin films with and without oxidant thermal annealing. a**, Film thickness and **b**, electrical conductivity variations. The red curves are films made with non-annealed oxidant and the black curves are those prepared with annealed oxidant. c, Optical extinction spectra for UV-treated PEDOT made by oxidants with (solid) and without thermal annealing (dashed).

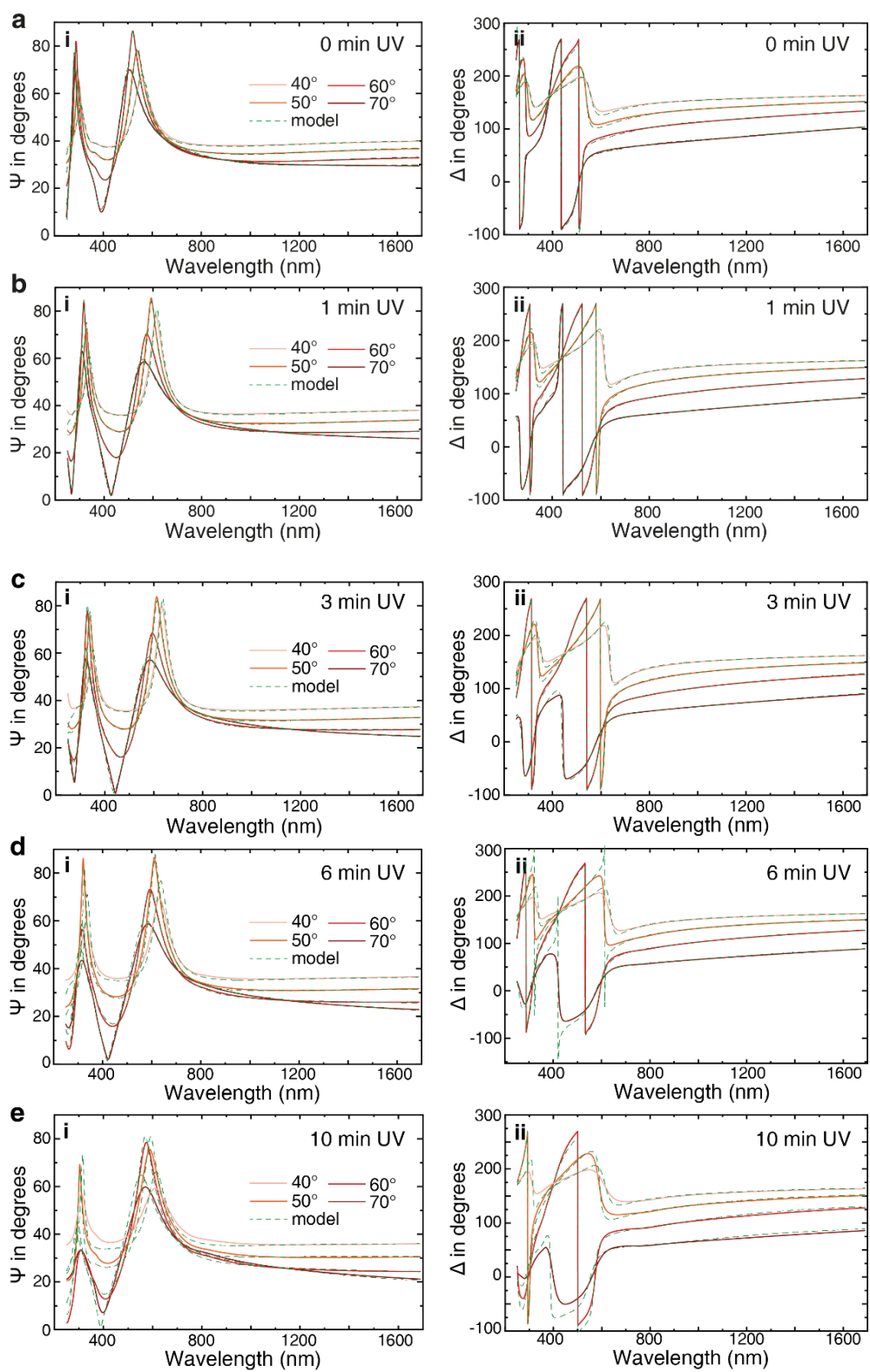

**Figure S27 | Spectroscopic ellipsometry spectra for PEDOT:Tos films using UV-exposed oxidants with post-baking.** Five UV exposure times are used: 0 min (**a**), 1 min (**b**), 3 min (**c**), 6 min (**d**), and 10 min (**e**). For each measurement, four incident angles are used: 40°, 50°, 60°, and 70°. The experimental measured data (**i** for ψ and **ii** for Δ) and model fits are plotted in solid and dashed curves. All the samples were deposited on silicon wafers coated with 100 nm Au films.

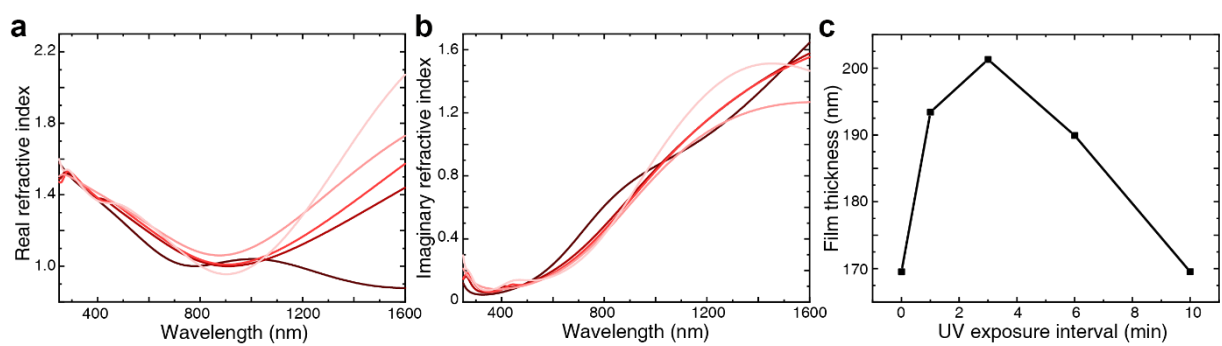

**Figure S28 | Properties of UV-treated PEDOT made with annealed oxidants. a**, In-plane real refractive index. **b**, In-plane imaginary refractive index. **c**, Thickness of the UV-patterned PEDOT determined by ellipsometry. UV exposure times of 0 min, 1 min, 3 min, 6 min, and 10 min were used.